\documentclass[11pt]{article}
\pdfoutput=1
\usepackage{jcappub,natbib,float,caption,comment,bm}
\usepackage{graphicx,epsfig}

\def\be{\begin{equation}}
\def\ee{\end{equation}}
\def\ba#1\ea{\begin{align}#1\end{align}}
\newcommand{\vs}{\nonumber\\}
	
\renewcommand{\v}[1]{\mathbf{#1}}
\newcommand{\vx}{\v{x}}
\newcommand{\vr}{\v{r}}	
\newcommand{\vk}{\v{k}}
\newcommand{\vq}{\v{q}}

\newcommand{\refeq}[1]{eq.~(\ref{eq:#1})}
\newcommand{\refeqs}[2]{eqs.~(\ref{eq:#1})--(\ref{eq:#2})}
\newcommand{\refEq}[1]{Eq.~(\ref{eq:#1})}
\newcommand{\reffig}[1]{figure~\ref{fig:#1}}
\newcommand{\refFig}[1]{Figure~\ref{fig:#1}}
\newcommand{\refsec}[1]{section~\ref{sec:#1}}
\newcommand{\refapp}[1]{appendix~\ref{app:#1}}
\newcommand{\reftab}[1]{table~\ref{tab:#1}}
\newcommand{\refTab}[1]{Table~\ref{tab:#1}}

\newcommand{\Om}{\Omega_m}

\newcommand{\Ob}{\Omega_b}

\renewcommand{\d}{\delta}

\newcommand{\cO}{\mathcal O}
\newcommand{\dr}{\delta({\bf r})}

\newcommand{\bd}{\bar\delta}
\newcommand{\iz}{i\zeta}
\newcommand{\se}{\sigma_8}
\newcommand{\sef}{\sigma_{8,\rm fid}}
\newcommand{\hMpc}{~h^{-1}~{\rm Mpc}}

\title{Position-dependent correlation function from the SDSS-III
Baryon Oscillation Spectroscopic Survey Data Release 10 CMASS Sample}
\author[a]{Chi-Ting Chiang,}
\author[a]{Christian Wagner,}
\author[b]{Ariel G. S{\'a}nchez,}
\author[a]{Fabian~Schmidt,}
\author[a,c]{and Eiichiro Komatsu}
\affiliation[a]{Max-Planck-Institut f\"ur Astrophysik, Karl-Schwarzschild-Str. 1, 85741 Garching, Germany}
\affiliation[b]{Max-Planck-Institut f\"ur Extraterrestrische Physik, Postfach 1312, Giessenbachstr., 85748 Garching, Germany}
\affiliation[c]{Kavli Institute for the Physics and Mathematics of the
Universe, Todai Institutes for Advanced Study, the University of Tokyo,
Kashiwa, Japan 277-8583 (Kavli IPMU, WPI)}
\emailAdd{ctchiang@mpa-garching.mpg.de}
\abstract{We report on the first measurement of the three-point function
with the {\it position-dependent correlation function} from the SDSS-III
Baryon Oscillation Spectroscopic Survey (BOSS) Data Release 10 CMASS sample.
This new observable measures the correlation between two-point functions
of galaxy pairs within different subvolumes, $\hat\xi(\vr,\vr_L)$, where
$\vr_L$ is the location of a subvolume, and the corresponding mean overdensities,
$\bar\delta(\vr_L)$. This correlation, which we call the ``integrated three-point function'',
$i\zeta(r)\equiv\langle\hat\xi(\vr,\vr_L)\bar\delta(\vr_L)\rangle$,
measures a three-point function of two short- and one long-wavelength modes,
and is generated by nonlinear gravitational evolution and possibly also by
the physics of inflation. The $i\zeta(r)$ measured from the BOSS data lies
within the scatter of those from the mock galaxy catalogs in redshift space,
yielding a ten-percent-level determination of the amplitude of $i\zeta(r)$.
The tree-level perturbation theory in redshift space predicts how this amplitude
depends on the linear and quadratic nonlinear galaxy bias parameters ($b_1$
and $b_2$), as well as on the amplitude and linear growth rate of matter
fluctuations ($\sigma_8$ and $f$). Combining $i\zeta(r)$ with the constraints
on $b_1\sigma_8$ and $f\sigma_8$ from the global two-point correlation
function and that on $\sigma_8$ from the weak lensing signal of BOSS
galaxies, we measure $b_2=0.41\pm0.41$ (68\% C.L.) assuming standard
perturbation theory at  the tree level and the local bias model.
}

\begin{document}
\maketitle
\flushbottom

\section{Introduction}
Mode coupling plays a fundamental role in cosmology. A long-wavelength scalar
density fluctuation modifies the formation of small-scale structure via gravitational
evolution (see \cite{bernardeau/etal:2001} for a review), and possibly also
through the physics of inflation. This effect of long-wavelength modes manifests
itself through a dependence of observables, e.g., the $n$-point statistics and
the halo mass function,  on the local long-wavelength overdensity, or equivalently,
the position in space (see \cite{chiang/etal:2014,wagner/etal:2014,wagner/etal:2015}
for the $n$-point statistics and \cite{cole/kaiser:1989,mo/white:1996} 
for the mass function). Measurements of spatially-varying, ``position-dependent''
observables capture the effects of mode coupling, and can be used to test our
understanding of gravity and the physics of inflation. A similar idea of measuring
the shift of the peak position of the baryonic acoustic oscillation in different
environments has been studied in ref.~\cite{roukema/etal:2014}.

In this paper, we focus on the position-dependent two-point function. Consider
a galaxy redshift survey. Instead of measuring the two-point function of
galaxy pairs within the entire survey volume, we divide the survey volume into many
subvolumes, within which we measure the two-point function of galaxy
pairs. These two-point functions vary spatially from subvolume to subvolume, and the
variation is correlated with the mean overdensities of the subvolumes
with respect to the entire survey volume. As we show later in detail, this correlation
measures an integral of the three-point function, which represents the
response of the small-scale clustering of galaxies (as measured by the
position-dependent two-point functions) to the long-wavelength density
perturbation (as measured by the mean overdensities of the subvolumes) \cite{chiang/etal:2014}.

Not only is the position-dependent correlation function conceptually straightforward
to interpret, but the computational requirement for measuring three-point statistics
is also largely alleviated. The usual three-point correlation function measurements
rely on finding particle triplets  with the naive algorithm scaling as $N_{\rm par}^3$
where $N_{\rm par}$ is the number of particles. Current galaxy redshift surveys contain
roughly a million galaxies, and we need 50 times as many random samples as the galaxies
for characterizing the survey window function accurately (see, e.g., \cite{chiang/etal:2013}).
Counting triplets thus becomes computationally challenging. Similarly, the measurement
of the three-point function in Fourier space, i.e., the bispectrum, requires counting
of all possible triangle configurations formed by different Fourier modes, which is also
computationally demanding. This explains why only few measurements of the three-point
function of the large-scale structure have been reported in the literature
\cite{scoccimarro/etal:2000,verde/etal:2001,kayo/etal:2004,nishimichi/etal:2006,
mcbride/etal:2010a,mcbride/etal:2010b,marin/etal:2013,gilmarin/etal:2014b,guo/etal:2014}.

The computational requirement is alleviated for the position-dependent
correlation function technique because we explore a subset of triplets
corresponding to the ``squeezed configurations'' of the three-point function;
namely, two short-wavelength modes correlated with one long-wavelength mode.
We only need to count particle pairs for measuring the two-point function
in subvolumes, which scales as $N_{\rm par}^2$. The scaling is further
improved because the number of particles in each subvolume decreases by
the number of the subvolumes $N_s$; hence, it scales as
$N_s(N_{\rm par}/N_s)^2=N_{\rm par}^2/N_s$.
The position-dependent correlation function technique is thus particularly
efficient for extracting the information of the squeezed-limit bispectrum,
which we shall demonstrate in \refapp{fisher}, while it is relatively
insensitive to the bispectrum in other configurations.

In this paper, we report on the first measurement of the three-point function
with the position-dependent correlation function from the SDSS-III Baryon
Oscillation Spectroscopic Survey Data Release 10 (hereafter BOSS DR10) CMASS
sample \cite{ahn/etal:2013,anderson/etal:2013}. We compare this measurement
with those from the PTHalos mock catalogs
\cite{scoccimarro/sheth:2001,manera/etal:2012,manera/etal:2014}.
While the mocks were designed to reproduce the global two-point function of
the BOSS DR10 CMASS sample, it is not guaranteed that they can reproduce the
three-point function as measured by the position-dependent correlation function.
We shall show that the position-dependent correlation functions from the real
data and the mocks are consistent with each other. Finally, we use tree-level
perturbation theory to predict the position-dependent correlation function as
a function of the galaxy bias parameters and the cosmological parameters, and
determine the quadratic nonlinear bias parameter of the BOSS DR10 CMASS sample
by combining the constraints from the position-dependent correlation function,
the global two-point function, and the weak lensing signal.

The rest of the paper is organized as follows. In \refsec{theory}, we define the
position-dependent correlation function and the integrated three-point function,
and describe the tree-level perturbation theory prediction for the integrated
three-point function in redshift space. In \refsec{mock} and \ref{sec:data},
we apply the position-dependent correlation function technique to the mocks
and the BOSS DR10 CMASS sample, respectively. The cosmological interpretation
of the measurements is given in \refsec{interpretation}. We conclude in \refsec{conclusion}.
In \refapp{gaussian}, we test our estimator using Gaussian realizations.
In \refapp{test}, we study the effects of using extended models of the
bispectrum with the effective $F_2$ and $G_2$ kernels and a tidal bias.
In \refapp{iz_zevolve}, we compare the mocks and BOSS DR10 CMASS samples in
different redshift bins.
In \refapp{fisher}, we use the Fisher matrix calculation to demonstrate
the information content of the position-dependent correlation function.
Throughout the paper we adopt the  cosmology of the
mocks as our fiducial cosmology, i.e., a flat $\Lambda$CDM cosmology
with $\Om=0.274$, $\Ob h^2=0.0224$, $h=0.7$, $\se=0.8$, and $n_s=0.95$.

%%%%%%%%%%%%%%%%%%%%%%%%%%%%%%%%%%%%%%%%%%%%%%%%%%%
\section{Position-dependent correlation function and the integrated
three-point function in redshift space}
\label{sec:theory}
\subsection{Position-dependent correlation function}
\label{sec:pos_dep_xi}

Consider a density fluctuation field, $\dr$, in a survey (or simulation) 
volume $V_r$. The mean overdensity of this volume vanishes by
construction, i.e.,  $\bd=\frac{1}{V_r}\int_{V_r}d^3r~\dr=0$.
The global two-point function is defined as
\be
 \xi(r)=\langle\delta(\vx)\delta(\vx+\vr)\rangle ~,
\ee
where we assume that $\dr$ is statistically homogeneous and isotropic,
so $\xi(r)$ depends only on the separation $r$. As the ensemble average
cannot be measured directly, we estimate the global two-point function as
\be
 \hat{\xi}(r)=\frac{1}{V_r} \int \frac{d^2\hat{\vr}}{4\pi}
 \int_{\vx,\vx+\vr\in V_r}d^3x~\d(\vr+\vx)\d(\vx)\,.
\label{eq:hat_xi_g}
\ee

The ensemble average of \refeq{hat_xi_g} is not equal to $\xi(r)$. Specifically,
\ba
 \langle\hat\xi(r)\rangle=\:&\frac{1}{V_r}\int\frac{d^2\hat{r}}{4\pi}
 \int_{\vx,\vx+\vr\in V_r}d^3x~\langle\d(\vr+\vx)\d(\vx)\rangle
 =\xi(r)\frac{1}{V_r}\int\frac{d^2\hat{r}}{4\pi}\int_{\vx,\vx+\vr\in V_r} d^3x~.
\label{eq:ensavg_hat_xi_g}
\ea
The second integral in \refeq{ensavg_hat_xi_g} is $V_r$ only if $\vr=0$, and
the fact that it departs from $V_r$ is due to the finite boundary of $V_r$.
We shall quantify this boundary effect later in \refeq{ensavg_hat_xi}.

We now identify a subvolume $V_L$ centered at $\vr_L$, and compute the mean
overdensity and the correlation function within $V_L$. The mean overdensity is
\be
 \bd(\vr_L)=\frac{1}{V_L}\int_{V_L}d^3r~\dr=\frac{1}{V_L}\int d^3r~\dr W(\vr-\vr_L) ~,
\ee
where $W(\vr)$ is the window function. Throughout this paper, we use a
cubic window function given by
\be
 W(\vr)=W_L(\vr)=\prod_{i=1}^3\:\theta(r_i), \quad
 \theta(r_i) = \left\{
 \begin{array}{cc}
  1, & |r_i|\le L/2, \\
  0,  & \mbox{otherwise}~,
 \end{array}\right.
\ee
where $L$ is the side length of $V_L$.
The results are not sensitive to the exact choice of the window
function, provided that the separation between galaxy pairs is much smaller
than $L$.
While $\bd=0$, $\bd(\vr_L)$ is non-zero in general. In other words, if $\bd(\vr_L)$
is positive (negative), then this subvolume is overdense (underdense) with respect
to the mean density in $V_r$.

Using the same window function, we define the position-dependent correlation
function in the subvolume $V_L$ centered at $\vr_L$ as
\ba
 \hat\xi(\vr,\vr_L)=\:&\frac{1}{V_L}\int\displaylimits_{\vx,\vr+\vx \in V_L}d^3x~\d(\vr+\vx)\d(\vx) \vs
 =\:&\frac{1}{V_L}\int d^3x~\d(\vr+\vx)\d(\vx)W_L(\vr+\vx-\vr_L)W_L(\vx-\vr_L) ~.
\ea
This is essentially an estimator for a local two-point function.
In this paper we shall consider only the angle-averaged position-dependent
correlation function (i.e.,~the monopole) defined by
\be
 \hat\xi(r,\vr_L)=\int\frac{d^2\hat{r}}{4\pi}~\hat\xi(\vr,\vr_L)
 =\frac{1}{V_L}\int\frac{d^2\hat{r}}{4\pi}\int d^3x~
 \d(\vr+\vx)\d(\vx)W_L(\vr+\vx-\vr_L)W_L(\vx-\vr_L) ~.
\label{eq:hat_xi}
\ee

Similarly to that of the global two-point function, the ensemble average
of \refeq{hat_xi} is not equal to $\xi(r)$.
Specifically,
\ba
 \langle\hat\xi(r,\vr_L)\rangle=\:&\frac{1}{V_L}\int\frac{d^2\hat{r}}{4\pi}\int d^3x~
 \langle\d(\vr+\vx)\d(\vx)\rangle W_L(\vr+\vx-\vr_L)W_L(\vx-\vr_L) \vs
 =\:&\xi(r)\frac{1}{V_L}\int\frac{d^2\hat{r}}{4\pi}\int d^3x'~
 W_L(\vr+\vx')W_L(\vx')\equiv\xi(r)f_{\rm bndry}(r) ~,
\label{eq:ensavg_hat_xi}
\ea
where $f_{\rm bndry}(r)$ is the boundary effect due to the finite size of
the subvolume. While $f_{\rm bndry}(r)=1$ for $r=0$, the boundary effect
becomes larger for larger separations. The boundary effect can be computed
by the five-dimensional integral in \refeq{ensavg_hat_xi}. Alternatively,
it can be evaluated by the ratio of the number of the random particle pairs
of a given separation in a finite volume to the expected random particle pairs
in the shell with the same separation in an infinite volume.
We have evaluated $f_{\rm bndry}(r)$ in both ways, and the results are in
an excellent agreement.

As the usual two-point function estimators based on pair counting (such as
Landy-Szalay estimator which will be discussed in \refsec{sub_quan}) or grid
counting (which will be discussed in \refapp{gaussian}) do not contain the
boundary effect, when we compare the measurements to the model which is
calculated based on \refeq{hat_xi}, we shall divide the model by $f_{\rm bndry}(r)$
to correct for the boundary effect.

\subsection{Integrated three-point function}
\label{sec:int_xi}
The correlation between $\hat\xi(r,\vr_L)$ and $\bd(\vr_L)$ is given by
\ba
 \langle\hat\xi(r,\vr_L)\bd(\vr_L)\rangle=\:&
 \frac{1}{V_L^2}\int\frac{d^2\hat{r}}{4\pi}\int d^3x_1\int d^3x_2
 ~\langle\d(\vr+\vx_1)\d(\vx_1)\d(\vx_2)\rangle \vs
 & ~~~~~~~~~~~~~~~~~~~~~~~~~~~~~~~~~~ \times
 W_L(\vr+\vx_1-\vr_L)W_L(\vx_1-\vr_L)W_L(\vx_2-\vr_L) \vs
 =\:&\frac{1}{V_L^2}\int\frac{d^2\hat{r}}{4\pi}\int d^3x_1\int d^3x_2
 ~\zeta(\vr+\vx_1+\vr_L,\vx_1+\vr_L,\vx_2+\vr_L) \vs
 & ~~~~~~~~~~~~~~~~~~~~~~~~~~~~~~~~~~ \times
 W_L(\vr+\vx_1)W_L(\vx_1)W_L(\vx_2) ~,
\label{eq:iz}
\ea
where $\zeta(\vr_1,\vr_2,\vr_3)\equiv\langle\d(\vr_1)\d(\vr_2)\d(\vr_3)\rangle$
is the three-point correlation function. Because of the assumption of homogeneity
and isotropy, the three-point function depends only on the separations $|\vr_i-\vr_j|$
for $i\neq j$, and so $\langle\hat\xi(r,\vr_L)\bd(\vr_L)\rangle$ is independent
of $\vr_L$. Furthermore, as the right-hand-side of \refeq{iz} is an integral
of the three-point function, we will refer to this quantity as the ``integrated
three-point function,'' $\iz(r)\equiv\langle\hat\xi(r,\vr_L)\bd(\vr_L)\rangle$.

$\iz(r)$ can be computed if $\zeta(\vr_1,\vr_2,\vr_3)$ is known. For example,
standard perturbation theory (SPT) with the local bias model at the tree level
in real space gives 
\be
 \iz(r)=b_1^3\iz_{\rm SPT}(r)+b_1^2b_2\iz_{b_2}(r)\,,
\ee
where  $\iz_{\rm SPT}$ and $\iz_{b_2}$ are given below. Here, $b_1$ and $b_2$
are the linear and quadratic (nonlinear) bias parameters, respectively. Because
of the high dimensionality of the integral, we use the Monte Carlo integration
routine in the GNU Scientific Library to numerically evaluate $\iz(r)$. The
first term, $\iz_{\rm SPT}$, is given by \cite{jing/borner:1996,barriga/gaztanaga:2001}
\ba
 \zeta_{\rm SPT}(\vr_1,\vr_2,\vr_3)=\:&\frac{10}{7}\xi_l(r_{12})\xi_l(r_{23})
 +\mu_{12,23}[\xi_l'(r_{12})\phi_l'(r_{23})+\xi_l'(r_{23})\phi_l'(r_{12})] \vs
 \:&+\frac{4}{7}\Bigg\lbrace-3\frac{\phi_l'(r_{12})\phi_l'(r_{23})}{r_{12}r_{13}}
 -\frac{\xi_l(r_{12})\phi_l'(r_{23})}{r_{23}}-\frac{\xi_l(r_{23})\phi_l'(r_{12})}{r_{12}} \vs
 &~~~~~~~~ +\mu_{12,23}^2\left[\xi_l(r_{12})+\frac{3\phi_l'(r_{12})}{r_{12}}\right]
 \left[\xi_l(r_{23})+\frac{3\phi_l'(r_{23})}{r_{23}}\right]\Bigg\rbrace \vs
 \:&+2~{\rm cyclic} ~,
\label{eq:zeta_spt}
\ea
where $r_{12}=|\vr_1-\vr_2|$, $\mu_{12,23}$ is the cosine between $\vr_{12}$
and $\vr_{23}$, $'$ refers to the spatial derivative, and
\be
 \xi_l(r)\equiv\int\frac{dk}{2\pi^2}~k^2P_l(k){\rm sinc}(kr)\\, ~~~~~~
 \phi_l(r)\equiv\int\frac{dk}{2\pi^2}~P_l(k){\rm sinc}(kr)\\,
\label{eq:xil_phil}
\ee
with $P_l(k)$ being the linear matter power spectrum, and ${\rm
sinc}(x)=\sin(x)/x$. The subscript $l$ denotes the quantities in the
linear regime.

The second term, $\iz_{b_2}$, is the nonlinear local bias three-point function. Since halos
(galaxies) are biased tracers of the underlying matter density field, the local
bias prescription yields the density field of the biased tracers as
$\d_h(\vr)=b_1\d_m(\vr)+\frac{b_2}{2}\d_m^2(\vr)+...$, where $b_1$ and $b_2$ are
the linear and nonlinear biases, respectively, and $\d_m(\vr)$ is the matter density field
\cite{fry/gaztanaga:1992}. The nonlinear bias three-point function is then
\be
 \zeta_{b_2}(\vr_1,\vr_2,\vr_3)=\xi_l(r_{12})\xi_l(r_{23})+2~{\rm cyclic} ~.
\label{eq:zeta_b2}
\ee

\begin{figure}[t]
\centering
\includegraphics[width=0.8\textwidth]{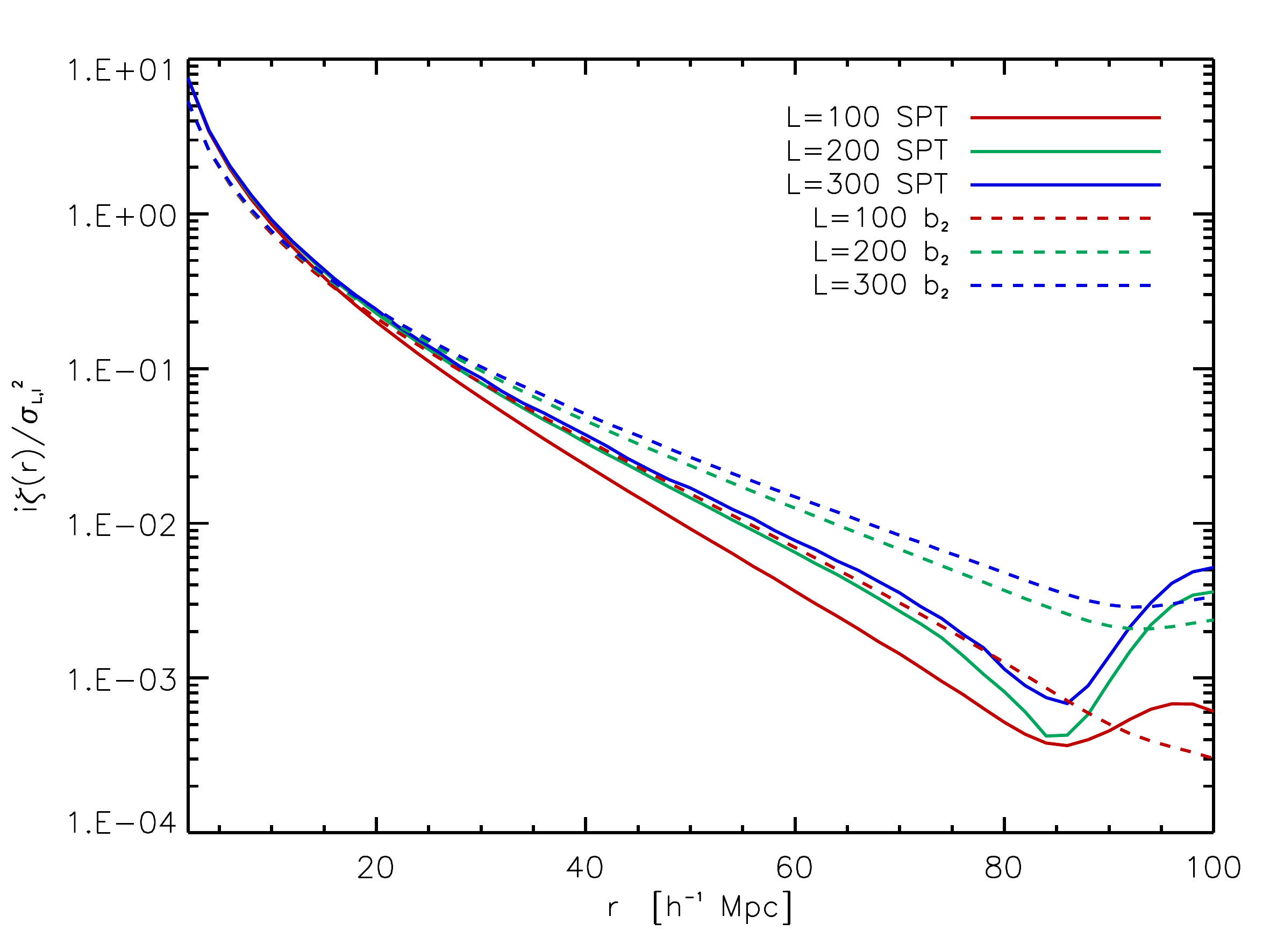}
\caption{Normalized $\iz_{\rm SPT}$ (solid) and $\iz_{b_2}$ (dashed) for $L=100\hMpc$ (red),
 $200\hMpc$ (green), and 300$\hMpc$ (blue) at $z=0$.}
\label{fig:iz_norm}
\end{figure}

\refFig{iz_norm} shows the scale-dependencies of $\iz_{\rm SPT}$ and $\iz_{b_2}$ at
$z=0$ with $P_l(k)$ computed by CLASS \cite{lesgourgues:2011}.
We normalize $\iz(r)$ by $\sigma_{L,l}^2$, where
\be
 \sigma_{L,l}^2\equiv\langle\bd_l(\vr_L)^2\rangle=\frac{1}{V_L^2}\int\frac{d^3k}{(2\pi)^3}~P_l(k)|W_L(\vk)|^2
\label{eq:sigmalL2}
\ee
is the variance of the linear density field in the subvolume $V_L$. The choice of
this normalization will become  clear in \refsec{squeezed} where we discuss $i\zeta$
in the squeezed limit, i.e., $r\ll L$. We find that the scale-dependencies of
$\iz_{\rm SPT}(r)$ and $\iz_{b_2}(r)$ are similar especially on small scales.
This is because the scale-dependence of the bispectrum in the squeezed limit is
(see e.g. the appendix of ref.~\cite{chiang/etal:2014})
\be
 B_{\rm SPT}\to\left[\frac{68}{21}-\frac13\frac{d\ln k^3P_l(k)}{d\ln k}\right]P_l(k)P_l(q) ~,~~
 B_{b_2}\to2P_l(k)P_l(q) ~,
\ee
where $k$ and $q$ are the short- and long-wavelength modes, respectively. For
a power-law power spectrum without features, the squeezed-limit $B_{\rm SPT}$
and $B_{b_2}$ have exactly the same scale dependence and cannot be distinguished.
This results in a significant residual degeneracy between $b_1$ and $b_2$, and will
be discussed in \refsec{interpretation}. When $r$ is small, $\iz(r)/\sigma_{L,l}^2$
becomes independent of the subvolume size. We derive this feature when we discuss
the squeezed limit in \refsec{squeezed}.

\subsection{Connection to the integrated bispectrum}
\label{sec:iz_to_ib}
Fourier transforming the density fields, the integrated three-point function
can be written as
\ba
 \iz(\vr) \:&=
\frac{1}{V_L^2}\int\frac{d^3q_1}{(2\pi)^3}\cdots\int\frac{d^3q_6}{(2\pi)^3}~(2\pi)^9
 \delta_D(\vq_1+\vq_2+\vq_3)\delta_D(\vq_1+\vq_2+\vq_4+\vq_5)\delta_D(\vq_3+\vq_6) \vs
 \:&~~~~~~~~~~~~~~~~~~\times B(\vq_1,\vq_2,\vq_3)W_L(\vq_4)W_L(\vq_5)W_L(\vq_6)
 e^{i[\vr\cdot(\vq_1+\vq_4)-\vr_L\cdot(\vq_4+\vq_5+\vq_6)]} \vs
 \:&=\int\frac{d^3k}{(2\pi)^3}~iB(\vk)e^{i\vr\cdot\vk} ~,
\label{eq:iz_ft_d}
\ea
where $B(\vq_1,\vq_2,\vq_3)$ is the bispectrum of the tracers, 
and
\be
 iB(\vk)\equiv\frac{1}{V_L^2}\int\frac{d^3q_1}{(2\pi)^3}\int\frac{d^3q_3}{(2\pi)^3}~
 B(\vk-\vq_1,-\vk+\vq_1+\vq_3,-\vq_3)W_L(\vq_1)W_L(-\vq_1-\vq_3)W_L(\vq_3)\,,
\label{eq:ib}
\ee
is the integrated bispectrum as defined in eq.~(2.7) of
ref.~\cite{chiang/etal:2014}. Eq.~\eqref{eq:iz_ft_d} shows that the integrated
three-point function is the Fourier transform of the integrated bispectrum.
Similarly, the angle-averaged integrated three-point function is related to the
angle-averaged integrated bispectrum, $iB(k)\equiv(4\pi)^{-1}\int{d^2\hat k}~iB(\vk)$, as
\ba
 \iz(r)=\int\frac{k^2dk}{2\pi^2}~iB(k)\,{\rm sinc}(kr)\,.
\label{eq:iz_ib_ang_avg}
\ea

\subsection{Squeezed limit}
\label{sec:squeezed}
In the squeezed limit, where the separation of the position-dependent correlation
function is much smaller than the size of the subvolume ($r\ll L$), the
integrated three-point function has a straightforward physical
interpretation \cite{chiang/etal:2014}.  In this case, the mean
density in the subvolume acts effectively as a constant ``background''
density. 
Consider the position-dependent correlation function, $\hat\xi(\vr,\vr_L)$,
measured in a subvolume with overdensity $\bd(\vr_L)$. If the overdensity
is small, we may Taylor expand $\hat\xi(\vr,\vr_L)$ in orders of $\bd$ as
\be
 \hat\xi(\vr,\vr_L) = \left.\xi(\vr)\right|_{\bd=0}+
 \left.\frac{d\xi(\vr)}{d\bd}\right|_{\bd=0}\bd+\cO(\bd^2) ~.
\label{eq:sep_uni}
\ee
The integrated three-point function in the squeezed limit is then, at leading order
in the variance $\langle\bar\delta^2\rangle$ (dropping $\bd=0$ in the subscript of
the derivative term for clarity), given by
\be
 \iz(\vr)=\langle\hat\xi(\vr,\vr_L)\bd(\vr_L)\rangle
 =\frac{d\xi(\vr)}{d\bd}\langle\bd^2\rangle+\cO(\bd^3) ~.
\label{eq:iz_squeezed_1}
\ee
As $\langle\bd^2\rangle=\sigma_L^2$\footnote{If $\bd=\bd_l$ then $\sigma_L^2=\sigma_{L,l}^2$.
But $\bd$ can in principle be nonlinear or the mean overdensity of the biased tracers,
so here we denote the variance to be $\sigma_L^2$.},
$\iz(\vr)$ normalized by $\sigma_L^2$ is $d\xi(\vr)/d\bd$ at leading order,
which is the linear response of the correlation function to the overdensity.
Note that in \refeq{iz_squeezed_1} there is no dependence on the subvolume
size apart from $\sigma_L^2$, as shown also by  the asymptotic behavior of
the solid lines in \reffig{iz_norm} for $r\to0$.

\begin{figure}[t]
\centering
\includegraphics[width=0.8\textwidth]{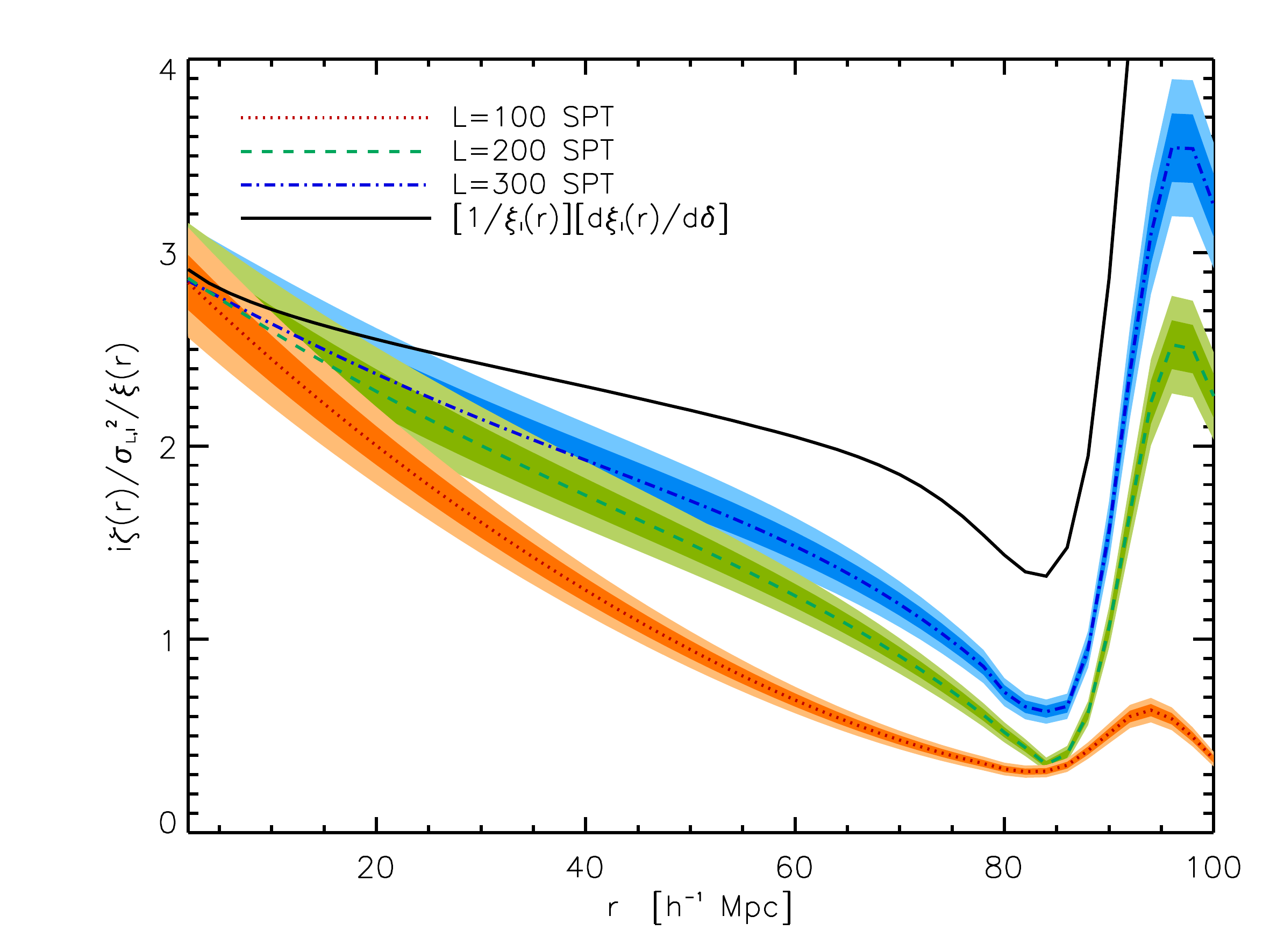}
\caption{The linear response function $[1/\xi_l(r)][d\xi_l(r)/d\bd]$ (black
solid) and the normalized $\iz_{\rm SPT}(r)$ for $L=100\hMpc$ (red dotted),
$200\hMpc$ (green dashed), and 300$\hMpc$ (blue dot-dashed). The light and
dark bands correspond to $\pm5\%$ and $\pm10\%$ of the predictions, respectively.} 
\label{fig:iz_norm_squ}
\end{figure}

As $i\zeta(r)$ is the Fourier transform of $iB(k)$, the response of the
correlation function, $d\xi(r)/d\bar\delta$, is also the Fourier
transform of the response of the power spectrum, $dP(k)/d\bar\delta$.
For example, we can calculate the response of the linear matter
correlation function, $d\xi_l(r)/d\bar\delta$, by Fourier transforming
$dP_l(k)/d\bar\delta=[68/21-(1/3)d\ln k^3P_l(k)/d\ln k]P_l(k)$
\cite{chiang/etal:2014}. In \reffig{iz_norm_squ}, we compare the normalized
$\iz_{\rm SPT}(r)$ with $d\xi_l(r)/d\bd$. Due to the large dynamic range of the
correlation function, we divide all the predictions by $\xi(r)$. As expected,
the smaller the subvolume size, the smaller the $r$ for $\iz_{\rm SPT}(r)$ to
be close to $[1/\xi_l(r)][d\xi_l(r)/d\bd]$, i.e., reaching the squeezed limit.
Specifically, for $100\hMpc$, $200\hMpc$, and $300\hMpc$ subvolumes,
the squeezed limit is reached to 10\% level at $r\sim10\hMpc$,
$18\hMpc$, and $25\hMpc$, respectively.

\subsection{Bispectrum in redshift space}
\label{sec:rsd}
To model $i\zeta(r)$ in redshift space, we need a model for the bispectrum in
redshift space. SPT in redshift space at the tree level predicts 
the halo bispectrum with local bias as \cite{scoccimarro/couchman/frieman:1999}
\be
 B_{z,{\rm tree-level}}(\vk_1,\vk_2,\vk_3)=2[Z_2(\vk_1,\vk_2)Z_1(\vk_1)Z_1(\vk_2)
 P_l(k_1)P_l(k_2)+2~{\rm cyclic}]\,,
\label{eq:treeredshift}
\ee
with
\ba
 Z_1(\vk_i)\:&=(b_1+f\mu_i^2)\,, \nonumber\\
 Z_2(\vk_1, \vk_2)\:&=b_1F_2(\vk_1,\vk_2)+f\mu^2G_2(\vk_1,\vk_2)+
 \frac{f\mu k}{2}\Big[\frac{\mu_1}{k_1}(b_1+f\mu_2^2)
 +\frac{\mu_2}{k_2}(b_1+f\mu_1^2)\Big]+\frac{b_2}{2} ~,
\ea
where $F_2$ and $G_2$ are the standard kernels of SPT
\cite{bernardeau/etal:2001}, $f=d\ln D/d\ln a$ is the logarithmic growth rate,
$\mu\equiv\hat{k}\cdot\hat{r}_{\rm los}$,
$\mu_i\equiv\hat{k}_i\cdot\hat{r}_{\rm los}$, and 
$\vk\equiv\vk_1+\vk_2$.

The integrated three-point function is the Fourier transform of the
integrated bispectrum. Thus, we can evaluate $\iz(r)$ by using
\refeq{treeredshift} in \refeq{ib} and averaging over the angle of
${\bf k}$ as in \refeq{iz_ib_ang_avg}. This operation requires a
nine-dimensional integral. On the other hand, if we have an expression
for the three-point function in configuration space, such as
\refeq{zeta_spt}, we can use \refeq{iz} to evaluate $\iz(r)$, which
requires an eight-dimensional integral. We do not always
have an analytical expression for the three-point function in
configuration space; thus, we in general need to perform the
nine-dimensional integral to obtain $\iz(r)$ from the bispectrum. 

Nevertheless, to check the precision of numerical integration, we compare the
results from the eight-dimensional integral in \refeq{iz} with
\refeq{zeta_spt}, and the nine-dimensional integral in \refeq{ib} and
\refeq{iz_ib_ang_avg} with $B({\bf k}_1,{\bf k}_2,{\bf
 k}_3)=2[F_2({\bf k}_1,{\bf k}_2)P_l(k_1)P_l(k_2)+2~\rm{cyclic}]$. As the
 latter gives a noisy result, we apply a Savitzky-Golay filter (with window size
9 and polynomial order 4) six times. We find that, on the scales of interest
($30\hMpc\le r\le78\hMpc$, which we will justify in \refsec{mock_r}),
both results are in agreement to  within 2\%. We repeat the same test
for  $\iz_{b_2}$ (\refeq{zeta_b2}), finding a similar result. As the
current uncertainty on the measured integrated correlation function
presented in this paper is of order 10\%, we 
conclude that our numerical integration yields sufficiently
precise results.

\subsection{Shot noise}
\label{sec:shot}
If the density field is traced by discrete particles, $\d_d(\vr)$, then the three-point
function contains a shot noise contribution given by
\ba
 \langle\d_d(\vr_1)\d_d(\vr_2)\d_d(\vr_3)\rangle\:&=\langle\d(\vr_1)\d(\vr_2)\d(\vr_3)\rangle \vs
 \:&+\left[\frac{\langle\d(\vr_1)\d(\vr_2)\rangle}{\bar n(\vr_3)}\d_D(\vr_1-\vr_3)+2~{\rm cyclic}\right]
 +\frac{\delta_D(\vr_1-\vr_2)\delta_D(\vr_1-\vr_3)}{\bar n(\vr_2)\bar n(\vr_3)} ~,
\label{eq:shot}
\ea
where $\bar n(r)$ is the mean number density of the discrete particles. The shot noise
can be safely neglected for the three-point function because it only contributes when
$\vr_1=\vr_2$, $\vr_1=\vr_3$, or $\vr_2=\vr_3$. On the other hand, the shot noise of
the integrated three-point function can be computed by inserting \refeq{shot} into
\refeq{iz}, which yields
\ba
 \iz_{\rm shot}(r)=\xi(r)\frac1{V_L^2}\int\frac{d^2\hat r}{4\pi}\int d^3x~
 \left[\frac1{\bar n(\vx+\vr+\vr_L)}+\frac1{\bar n(\vx+\vr_L)}\right]W_L(\vx+\vr)W_L(\vx) ~,
\ea
where we have assumed $r\neq0$. 
If we further assume that the mean number
density is constant, then the shot noise of the integrated three-point function can be
simplified as
\be
 \iz_{\rm shot}(r)=2\xi(r)\frac1{V_L\bar n}f_{\rm bndry}(r) ~.
\ee
For the measurements of PTHalos mock catalogs and the BOSS DR10 CMASS sample, the shot
noise is subdominant (less than 7\% of the total signal on the scales of interest).

%%%%%%%%%%%%%%%%%%%%%%%%%%%%%%%%%%%%%%%%%%%%%%%%%%%%%%%%%%%%%%%%%%%%%%%%%%%%

\section{Application to PTHalos mock catalogs}
\label{sec:mock}
Before we apply the position-dependent correlation function technique to
the real data, we apply it to the 600 PTHalos mock galaxy catalogs  of the
BOSS DR10 CMASS sample in the North Galactic Cap (NGC). From now on, we
refer to the real and mock BOSS DR10 CMASS samples as the ``observations''
and ``mocks'', respectively. 

We use the redshift range of $0.43<z<0.7$, and each realization of mocks
contains roughly 400,000 galaxies. We convert the positions of galaxies
in RA, DEC, and redshift to comoving distances using the cosmological
parameters of the mocks. The mocks have the same observational conditions
as the observations, and we correct the observational systematics by weighting
each galaxy differently. Specifically, we upweight a galaxy if its nearest
neighbor has a redshift failure ($w_{\rm zf}$) or a missing redshift due
to a close pair ($w_{\rm cp}$). We further apply weights to correct for
the correlation between the number density of the observed galaxies and
stellar density ($w_{\rm star}$) and seeing ($w_{\rm see}$).
We apply the same weights as done in the analyses of the BOSS collaboration,
namely FKP weighting, $w_{\rm FKP}=[1+P_w\bar n(z){\rm comp}]^{-1}$
\cite{feldman/kaiser/peacock:1993}, where $P_w=20000~h^{-3}~{\rm Mpc}^3$,
and $\bar n(z)$ and ``${\rm comp}$'' are the expected galaxy number density
and the survey completeness, respectively, provided in the catalogs.
Therefore, each galaxy is weighted by
$w_{\rm BOSS}=(w_{cp}+w_{zf}-1)w_{\rm star}w_{\rm see}w_{\rm FKP}$.

In this section, we present measurements from mocks in real space in
\refsec{mock_r} and redshift space in \refsec{mock_z}. The application
to the CMASS DR10 sample is the subject of \refsec{data}.

\subsection{Dividing the subvolumes}
\label{sec:division}
We use SDSSPix\footnote{SDSSPix: \url{http://dls.physics.ucdavis.edu/~scranton/SDSSPix}}
to pixelize the DR10 survey area. In short, at the lowest resolution (res=1)
SDSSPix divides the sphere equally into $n_x=36$ longitudinal slices across
the hemisphere (at equator each slice is 10 degrees wide), and each slice is
divided into $n_y=13$ pieces along constant latitudes with equal area. Thus,
for res=1 there are $n_x\times n_y=468$ pixels. In general the total number
of pixels is $n_x'\times n_y'=({\rm res}~n_x)\times({\rm res}~n_y)=(\rm res)^2\times468$,
and in this paper we shall set res=1024. After the pixelization, the $i^{\rm th}$
object (a galaxy or a random sample) has the pixel number $(i_x,i_y)$.

\begin{figure}[t]
\centering
\includegraphics[width=0.495\textwidth]{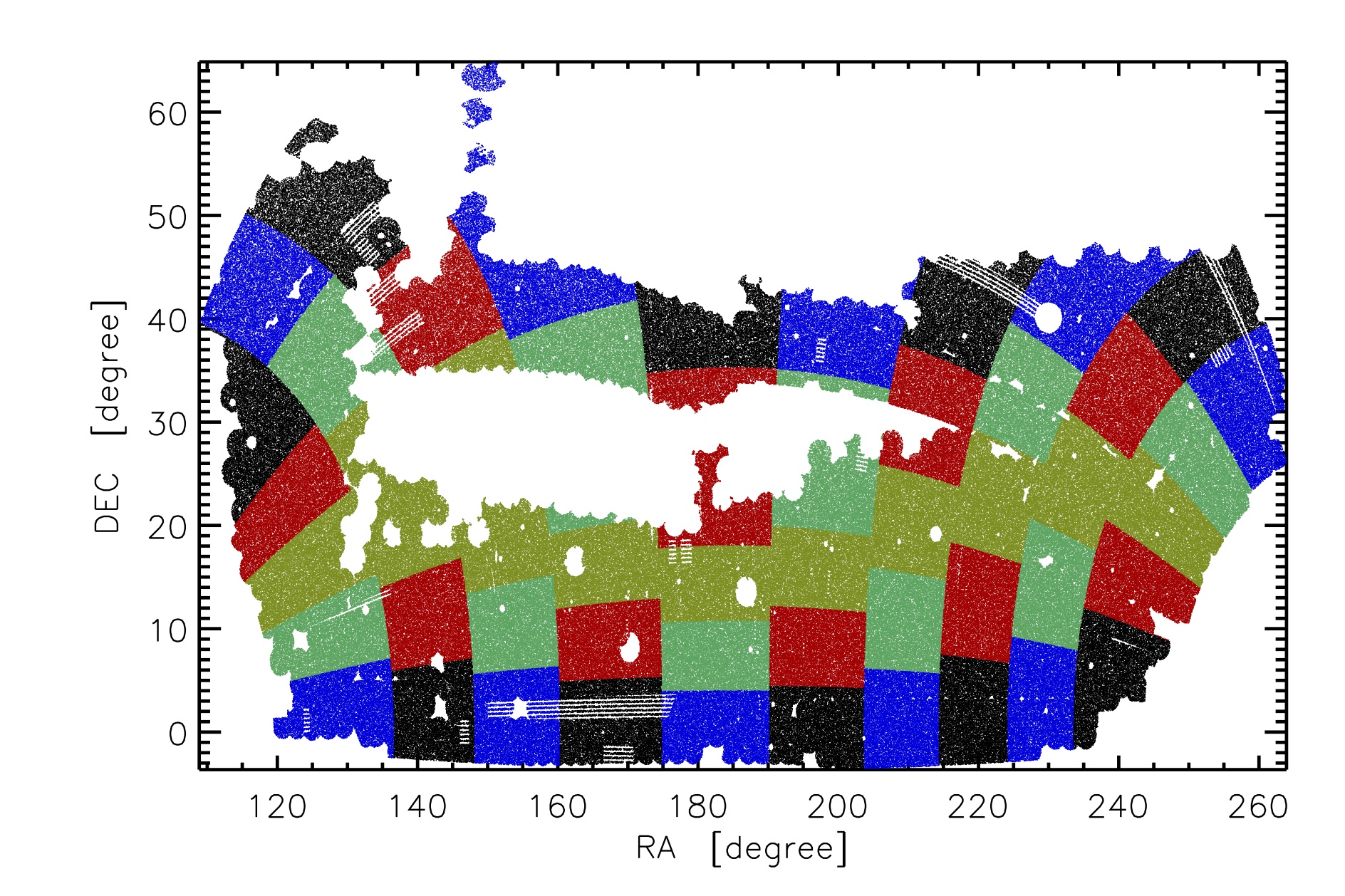}
\includegraphics[width=0.495\textwidth]{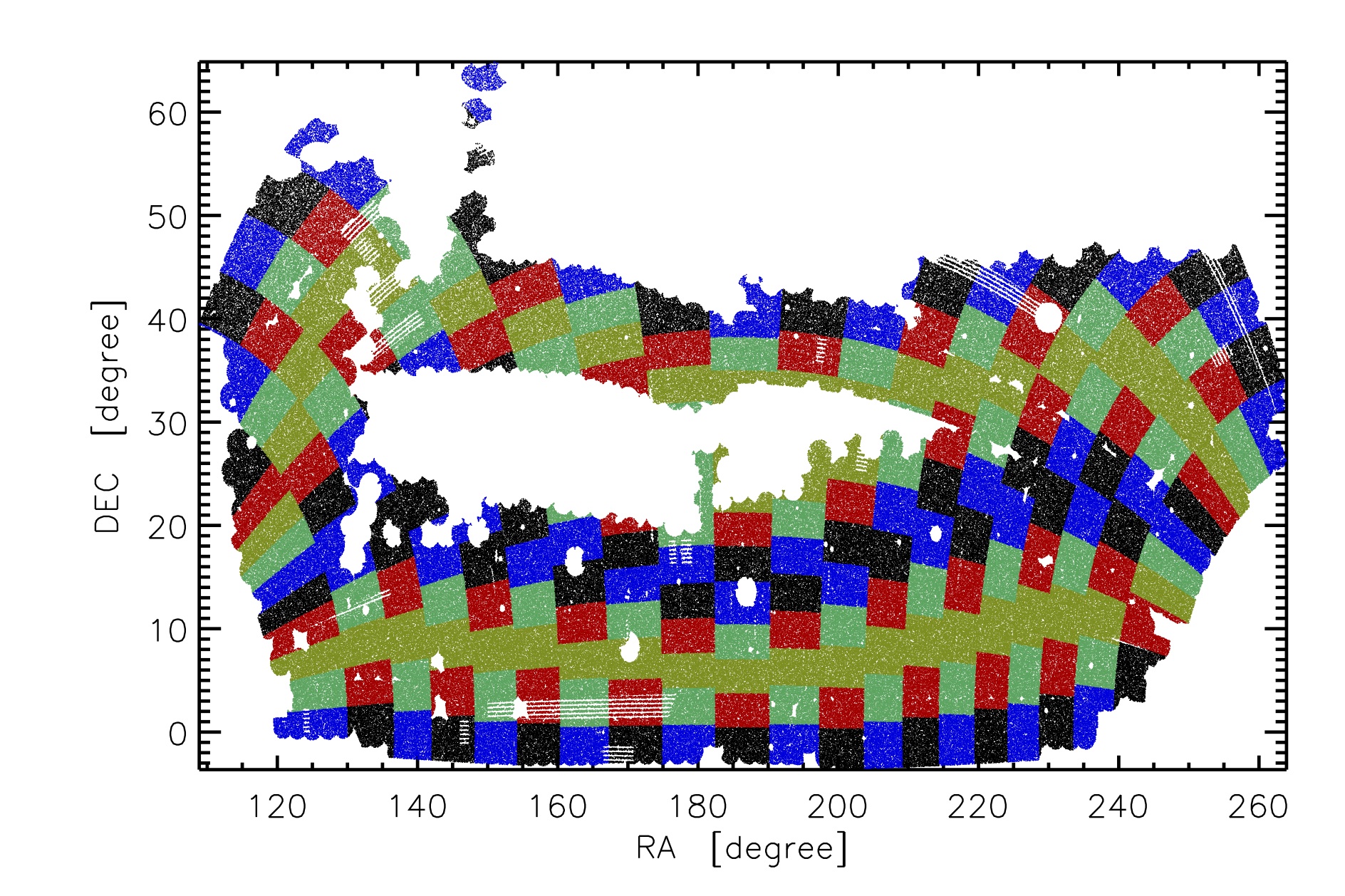}
\caption{Division of random samples into subvolumes with two resolutions
 in the RA-DEC plane. Each colored pattern extends over the redshift direction.}
\label{fig:ran_div}
\end{figure}

We use two different subvolume sizes. To cut the irregular survey volume into
subvolumes with roughly the same size, we first divide the random samples at
all redshifts into 10 and 20 slices across longitudes with similar numbers of
random samples; we then divide the random samples in each slice into 5 and 10
segments across latitudes with similar numbers of random samples. \refFig{ran_div}
shows the two resolutions of our subvolumes before the redshift cuts. (Note that
this resolution is different from the resolution of SDSSPix, which we always set
to res=1024.) Each colored pattern extends over the redshift direction. Finally,
we divide the two resolutions into three ($z_{\rm cut}=0.5108$, 0.5717) and five
($z_{\rm cut}=0.48710$, 0.52235, 0.55825, 0.60435) redshift bins.

As a result, there are 150 and 1000 subvolumes for the low and high resolution
configurations, respectively. The sizes of the subvolumes are  approximately
$V_L^{1/3}=220\hMpc$ and $120\hMpc$, respectively\footnote{The shapes of the
subvolumes are not exactly cubes. For example, for the high resolution, the ratios
of square root of the area to the depth, $\sqrt{L_xL_y}/L_z$, are roughly 0.78,
1.42, 1.51, 1.28, and 0.71, from the lowest to the highest redshift bins. The
results are not sensitive to the exact shape of the subvolumes, as long as the
separation of the position-dependent correlation function that we are interested
in is sufficiently smaller than $L_x$, $L_y$, and $L_z$.}. The fractional differences
between the numbers of the random samples in  subvolumes for the low and high
resolutions are within $^{+0.68\%}_{-0.58\%}$ and $^{+1.89\%}_{-1.83\%}$,
respectively. Since the number of random samples represents the effective volume,
all subvolumes at a given resolution have similar effective volumes. We assign
galaxies into subvolumes following the division of random samples.

\subsection{Estimators in the subvolumes}
\label{sec:sub_quan}
In the $i^{\rm th}$ subvolume, we measure the mean overdensity with respect
to the entire NGC, $\bd_i$, and the position-dependent correlation function,
$\hat\xi_i(r)$. The mean overdensity is estimated by comparing the total weighted
galaxies to the expected number density given by the random samples, i.e.,
\be
 \bd_i=\frac{1}{\alpha}\frac{w_{g,i}}{w_{r,i}}-1\,, ~~~~~
 \alpha\equiv\frac{\sum_{i=1}^{N_s}w_{g,i}}{\sum_{i=1}^{N_s}w_{r,i}}
 =\frac{w_{g,{\rm tot}}}{w_{r,{\rm tot}}} ~,
\ee
where $w_{g,i}$ and $w_{r,i}$ are the total weights ($w_{\rm BOSS}$) of galaxies
and random samples in the $i^{\rm th}$ subvolume, respectively, and $N_s$ is the
number of subvolumes. 

We use the Landy-Szalay estimator \cite{landy/szalay:1993} to estimate
the position-dependent correlation function as
\be
 \hat\xi_{{\rm LS},i}(r,\mu)=\frac{DD_i(r,\mu)}{RR_i(r,\mu)}
\left( \frac{[\sum_r w_{r,i}]^2-\sum_r w_{r,i}^2}{[\sum_g w_{g,i}]^2-\sum_g w_{g,i}^2}\right)
 -\frac{DR_i(r,\mu)}{RR_i(r,\mu)}\frac{([\sum_r w_{r,i}]^2-\sum_r w_{r,i}^2)}{\sum_g w_{g,i} \sum_r w_{r,i}}+1 ~,
\label{eq:ls_xi_est}
\ee
where $DD_i(r,\mu)$, $DR_i(r,\mu)$, and $RR_i(r,\mu)$ are the weighted numbers
of galaxy-galaxy, galaxy-random, and random-random pairs within the $i^{\rm th}$
subvolume, respectively, and $\mu$ is the cosine between the line-of-sight vector
and the vector connecting galaxy pairs ($\vr_1-\vr_2$). The summations such as
$\sum_{r}w_{r,i}$ and $\sum_{g}w_{g,i}$ denote the sum over all the random samples
and galaxies within the $i^{\rm th}$ subvolume, respectively. The angular average
correlation function is then $\hat\xi_{{\rm LS},i}(r)=\int_0^1d\mu~\hat\xi_{{\rm LS},i}(r,\mu)$. 

\refEq{ls_xi_est} estimates the correlation function assuming that the density
fluctuation is measured relative to the ${\it local}$ mean. However, the position-dependent
correlation function defined in \refsec{theory} uses the density fluctuation 
relative to the {\it global} mean. These two fluctuations can be related by
$\d_{\rm global}=(1+\bd)\d_{\rm local}+\bd$ with $\bd=\bar n_{\rm local}/\bar n_{\rm global}-1$.
Thus, the position-dependent correlation function, $\hat\xi_i(r)$, is related
to the Landy-Szalay estimator as
\be
 \hat\xi_i(r)=(1+\bd_i)^2\hat\xi_{{\rm LS},i}(r)+\bd_i^2 ~.
\label{eq:xi_corr}
\ee

To compute the average quantities over all subvolumes, we weight by $w_{r,i}$
in the corresponding subvolume. For example, for a given variable $g_i$ in the
$i^{\rm th}$ subvolume, the average over all subvolumes, $\bar g$, is defined by
\be
 \bar g=\frac{1}{w_{r,{\rm tot}}}\sum_{i=1}^{N_s}g_iw_{r,i} ~.
\label{eq:ens_avg}
\ee
Since the number of random samples in each subvolume represents the effective
volume, the average quantities are effective-volume weighted. \refEq{ens_avg}
assures that the mean of the individual subvolume overdensities is zero,
\be
 \bd=\frac{1}{w_{r,{\rm tot}}}\sum_{i=1}^{N_s}\bd_iw_{r,i}
 =\frac{1}{w_{r,{\rm tot}}}\sum_{i=1}^{N_s}\left[\frac{1}{\alpha}w_{g,i}-w_{r,i}\right]
 =\frac{\alpha}{\alpha}-1=0 ~.
\ee
We also confirm that $\bar{\hat{\xi}}(r)$ from \refeq{xi_corr} agrees with the
two-point function of all galaxies in the entire survey, on scales smaller than
the subvolume size.
integrated three-point function in the subvolume of size $L$ as
\be
 \iz(r)=\frac{1}{w_{r,{\rm tot}}}\sum_{i=1}^{N_s}\left[\hat\xi_i(r)\bd_i
 -2\bar{\hat\xi}(r)\frac{(1+\alpha)}{\alpha}\frac{\sum_rw^2_{r,i}}{\sum_r\bar n_{r,i}{\rm comp}_{r,i}w^2_{r,i}}
 \left(\sum_r\frac{1}{\bar n_{r,i}{\rm comp}_{r,i}}\right)^{-1}\right]w_{r,i} ~,
\ee
where the second term in the parentheses is the shot noise contribution, and
$\bar n_{r,i}$ and ${\rm comp}_{r,i}$ are the expected galaxy number density
and the survey completeness, respectively, of the random samples. Similarly,
we estimate the shot-noise-corrected variance of the fluctuations in the
subvolumes of size $L$ as
\be
 \sigma_L^2=\frac{1}{w_{r,{\rm tot}}}\sum_{i=1}^{N_s}
 \left[\bd_i^2-\frac{(1+\alpha)}{\alpha}\frac{\sum_rw^2_{r,i}}
 {\sum_r\bar n_{r,i}{\rm comp}_{r,i}w^2_{r,i}}
 \left(\sum_r\frac{1}{\bar n_{r,i}{\rm comp}_{r,i}}\right)^{-1}\right]w_{r,i} ~,
\ee
where the second term in the parentheses is the shot noise contribution.
We find that the shot noise is subdominant (less than 10\%) in both $\iz(r)$
and $\sigma_L^2$.

\subsection{Measurements in real space}
\label{sec:mock_r}
\begin{figure}[t]
\centering
\includegraphics[width=0.495\textwidth]{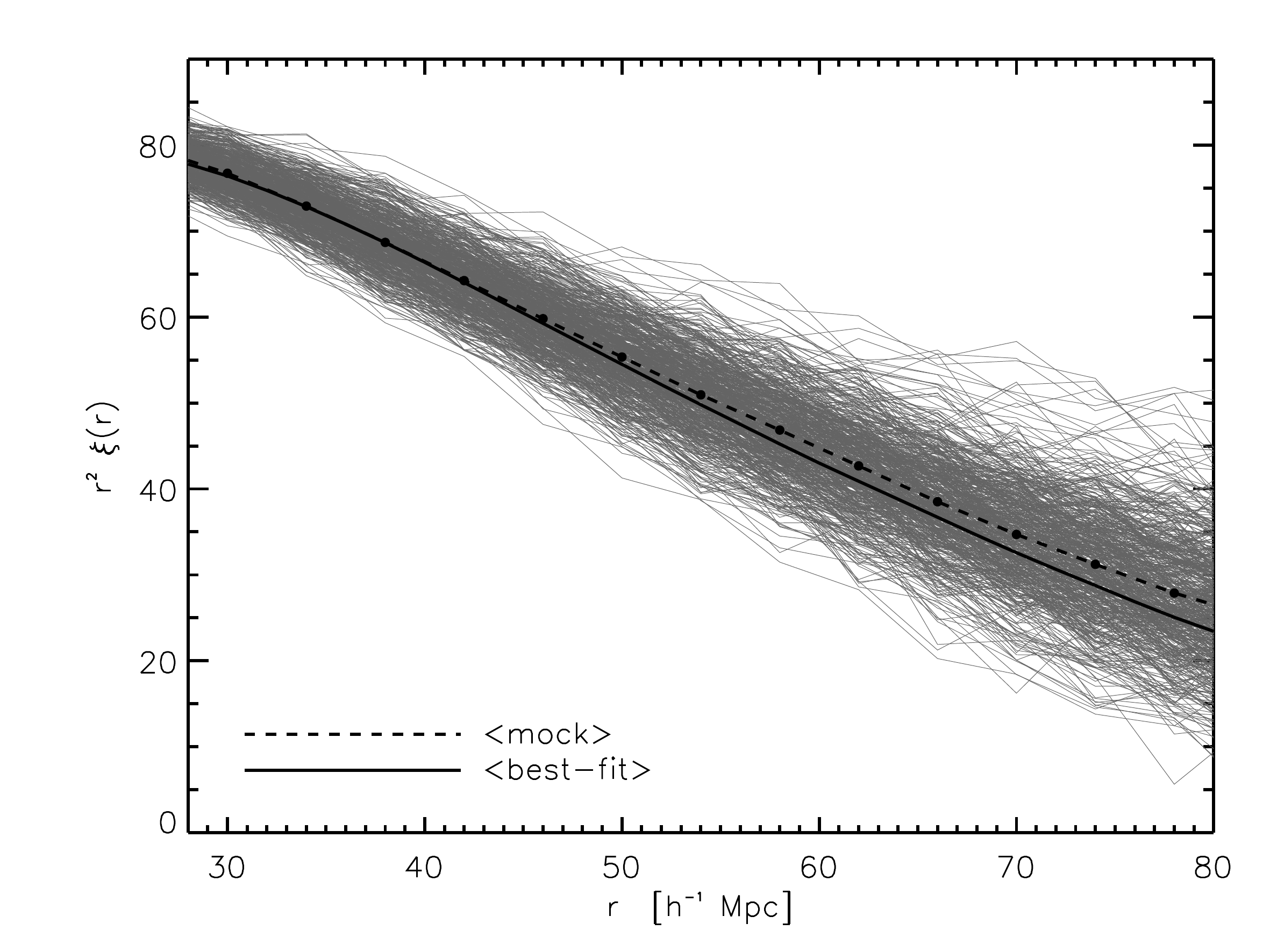}
\includegraphics[width=0.495\textwidth]{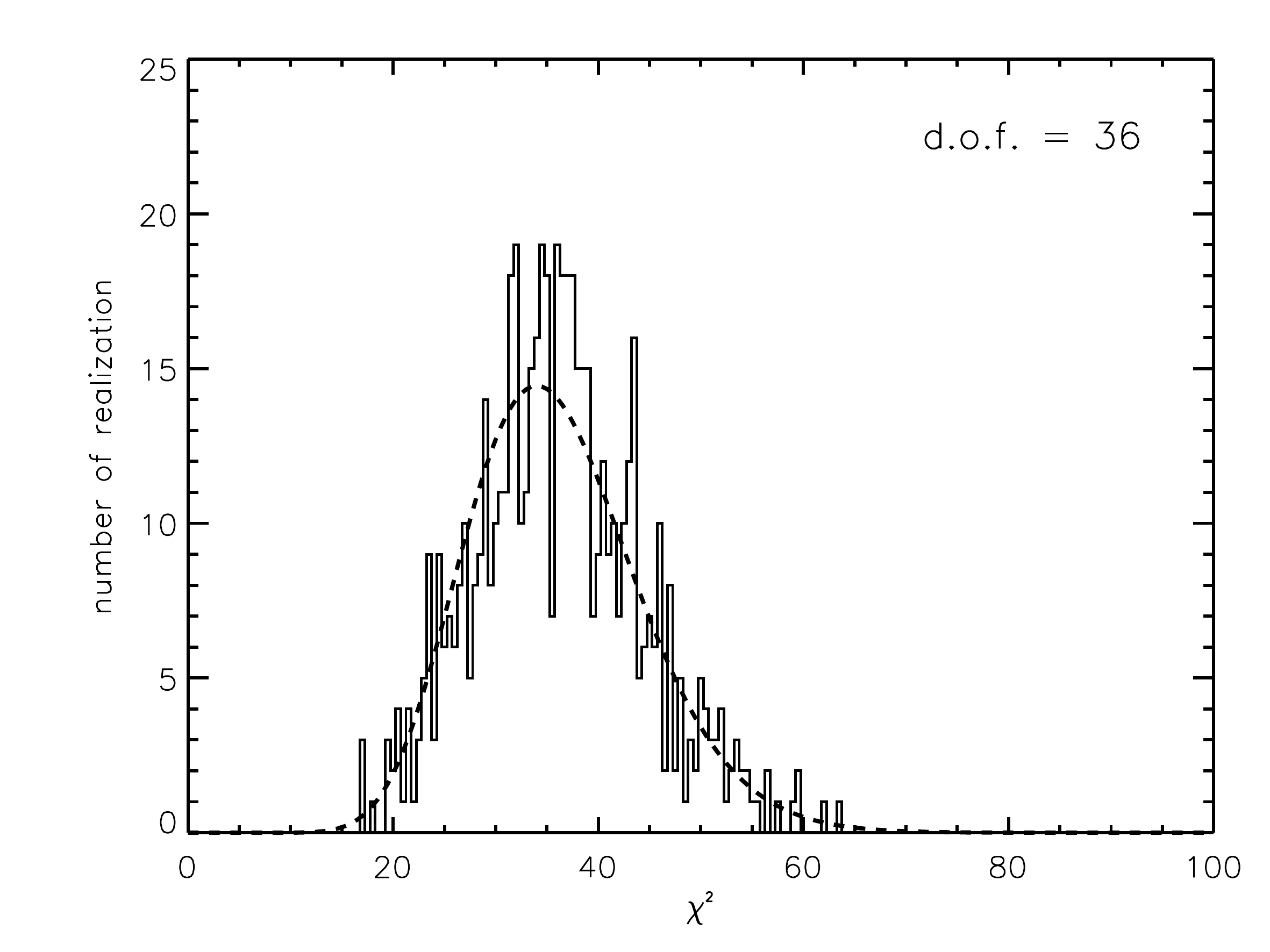}
\includegraphics[width=0.495\textwidth]{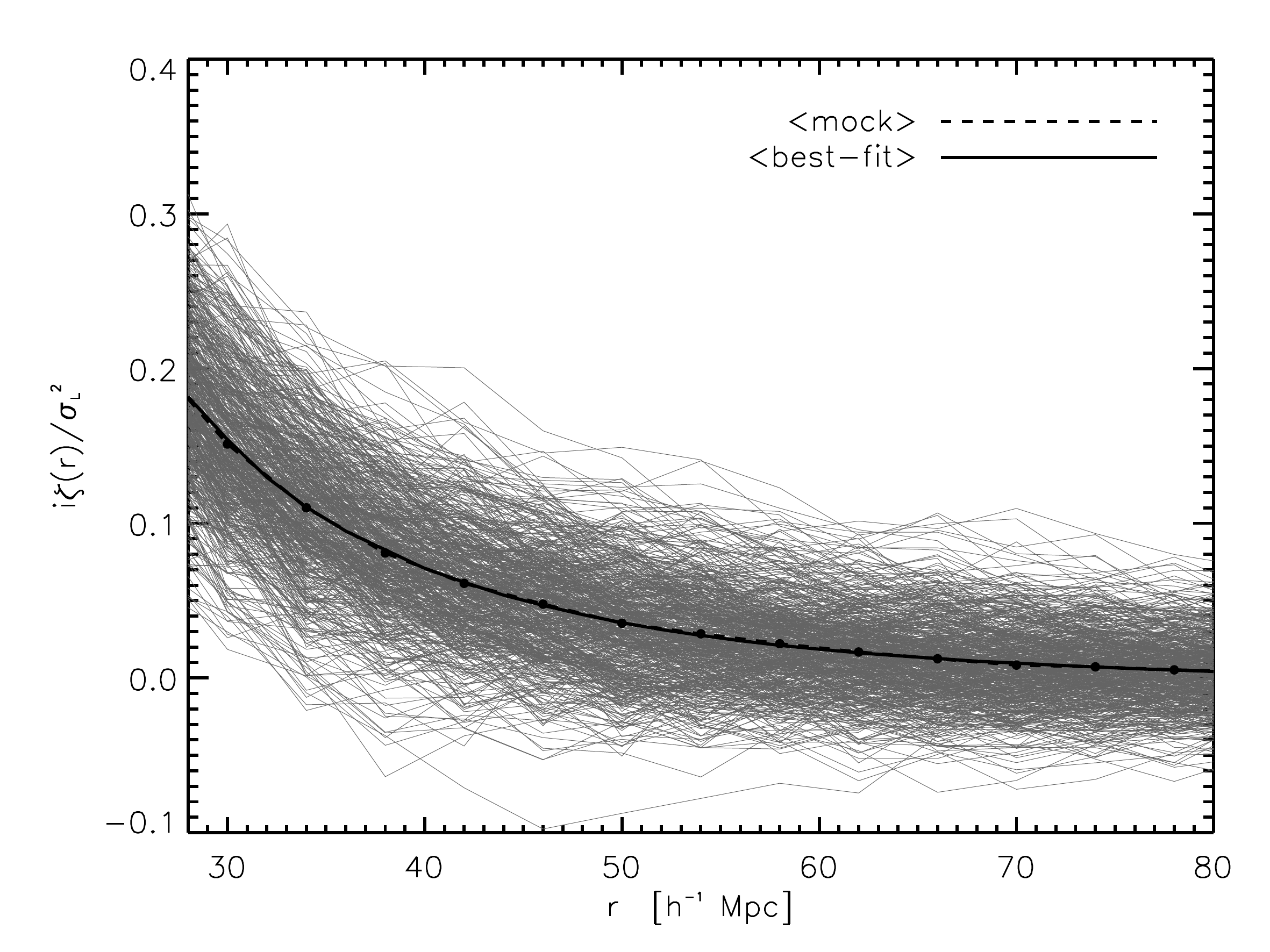}
\includegraphics[width=0.495\textwidth]{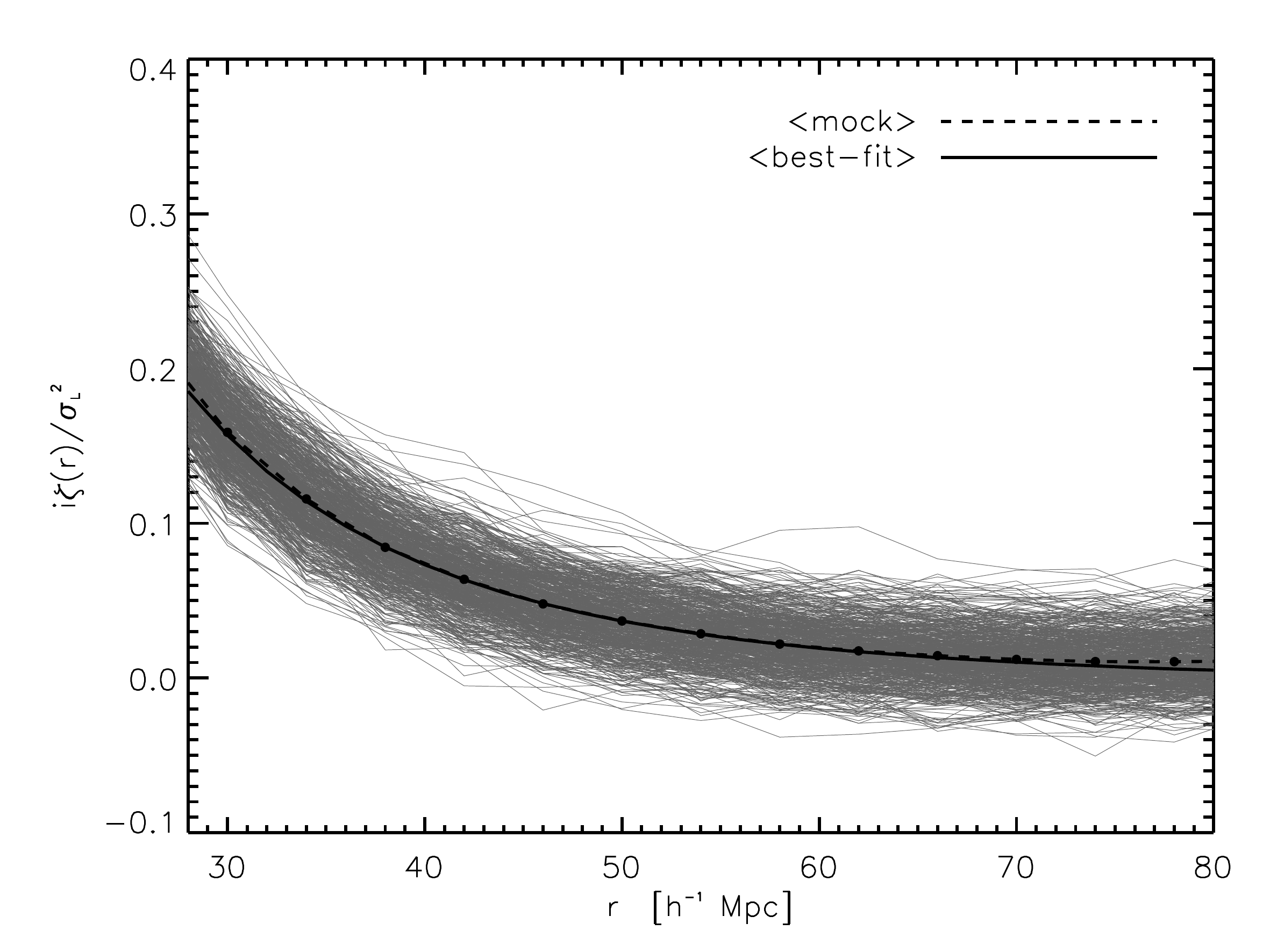}
\caption{(Top left) $\xi(r)$ of the mocks in real space. The gray lines show
individual realizations, while the dashed line shows the mean. The black solid
line shows the best-fitting model. (Top right) $\chi^2$-histogram of the 600
mocks jointly fitting the models to $\xi(r)$ and $\iz(r)/\sigma_L^2$ in real
space. The dashed line shows the $\chi^2$-distribution with d.o.f.=36. (Bottom
left) $\iz(r)/\sigma_L^2$ of the mocks in real space for $220\hMpc$ subvolumes.
(Bottom right) Same as the bottom left panel, but for $120\hMpc$ subvolumes.}
\label{fig:mock_r}
\end{figure}

\refFig{mock_r} shows the measurements of the two-point function
$\xi(r)$ from the entire survey (top left) and the normalized integrated
three-point functions (bottom panels), $\iz(r)/\sigma_L^2$,
%\be
% \frac{\iz(r)}{\sigma_L^2}=\left(\frac{1}{w_{r,{\rm tot}}}\sum_{i=1}^{N_s}
% \left[\hat\xi_i(r)\bd_iw_{r,i}\right]\right)\frac{1}{\sigma_L^2} ~,
%\ee
for the subvolumes of two sizes ($220\hMpc$ in the bottom-left and $120\hMpc$
in the bottom-right panels). The gray lines show individual realizations, while
the dashed lines show the mean.

We now fit models of $\xi(r)$ and $\iz(r)/\sigma_L^2$ to the measurements in
$30\hMpc\le r\le78\hMpc$. We choose this fitting range because there are less
galaxy pairs at larger separations due to the subvolume size, and the nonlinear
effect becomes too large for our SPT predictions to be applicable at smaller
separations. For the two-point function, we take the Fourier transform of
\cite{crocce/scoccimarro:2008}
\be
 P_g(k)=b_1^2[P_l(k)e^{-k^2\sigma_v^2}+A_{\rm MC}P_{\rm MC}(k)]~,
\ee
where $b_1$ is the linear bias, $P_l(k)$ is the linear power spectrum,
$A_{\rm MC}$ is the mode coupling constant, and
\be
 P_{\rm MC}(k)=2\int\frac{d^3q}{(2\pi)^3}~P_l(q)P_l(|\vk-\vq|)[F_2(\vq,\vk-\vq)]^2 ~.
\label{eq:xi_model}
\ee
Hence, $\xi_g(r)=b_1^2[\xi_{l,\sigma_v}(r)+A_{\rm MC}\xi_{\rm MC}(r)]$ with
\be
 \xi_{l,\sigma_v}(r)=\int\frac{d^3k}{(2\pi)^3}~P_l(k)e^{-k^2\sigma_v^2}e^{i\vk\cdot\vr}\,, ~~~~~
 \xi_{\rm MC}(r)=\int\frac{d^3k}{(2\pi)^3}~P_{\rm MC}(k)e^{i\vk\cdot\vr} \ .
\label{eq:xi_model_2}
\ee
We use a fixed value of $\sigma_v^2=20.644~h^{-2}~{\rm Mpc}^2$. Varying it has
only small effect on the other fitted parameters. For the integrated three-point
function, we use the SPT calculation
\be
 \frac{\iz_g(r)}{\sigma_L^2}
 =\frac{b_1\iz_{\rm SPT}(r)+b_2\iz_{b_2}(r)}{\sigma_{L,l}^2}\frac1{f_{\rm bndry}(r)} ~,
\label{eq:iz_model}
\ee
where $\iz_{\rm SPT}(r)$ and $\iz_{b_2}(r)$ are computed from \refeq{iz}
with eqs.~\eqref{eq:zeta_spt} and \eqref{eq:zeta_b2}, respectively, and
$\sigma_{L,l}^2$ is computed from \refeq{sigmalL2}, using the subvolume
sizes of $L=220$ and $120\hMpc$ and the redshift of $z=0.57$. Note that
the size of the subvolumes affects the values of $\sigma_{L,l}^2$. We
determine $L$ by first measuring $b_1^2$ using the real-space two-point
function of the entire survey, and then find $L$ such that $b_1^2\sigma_{L,l}^2=\sigma_L^2$
assuming the cubic top-hat window function\footnote{In principle, the
shape of the window function also affects $\sigma_{L,l}^2$, but we ignore
this small effect.}. We find that these values ($L=220$ and $120\hMpc$)
agree well with the cubic root of the total survey volume divided by the
number of subvolumes, to within a few percent.

We fit the models to $\xi(r)$ and $\iz(r)/\sigma_L^2$ of both subvolumes
simultaneously by minimizing
\be
 \chi^2=\sum_{ij}C^{-1}_{ij}(D_i-M_i)(D_j-M_j)\,,
\label{eq:chi2}
\ee
where $C^{-1}$ is the inverse covariance matrix computed from the 600 mocks,
$D_i$ and $M_i$ are the data and the model in the $i^{\rm th}$ bin, respectively.
The models contain three fitting parameters $b_1$, $b_2$, and $A_{\rm MC}$.

The models computed with the mean of the best-fitting parameters of 600 mocks
are shown as the black solid lines in \reffig{mock_r}.
The best-fitting parameters are $b_1=1.971\pm0.076$, $b_2=0.58\pm0.31$, and
$A_{\rm MC}=1.44\pm0.93$, where the error bars are 1-$\sigma$ standard deviations.
The agreement between the models and the mocks is good, with a difference much
smaller than the scatter among 600 mocks. Upon scrutinizing, the difference in
$\xi(r)$ is larger for larger separations because the fit is dominated by the
small separations with smaller error bars. On the other hand, for $\iz(r)/\sigma_L^2$
the agreement is good for both sizes of subvolumes at all scales of interest.
This indicates that the SPT calculation is sufficient to capture the three-point
function of the mocks in real space.

The data points in \reffig{mock_r} are highly correlated. To quantify the quality
of the fit, we compute the $\chi^2$-histogram from 600 mocks, and compare it with
the $\chi^2$-distribution with the corresponding degrees of freedom (d.o.f.). There
are 13 fitting points for each measurement ($\xi(r)$ and two sizes of subvolumes
for $\iz(r)/\sigma_L^2$) and three fitting parameters, so d.o.f.=36. The top right
panel of \reffig{mock_r} shows the $\chi^2$-histogram. The dashed line shows the
$\chi^2$-distribution with d.o.f.=36. The agreement is good, and we conclude that
our models well describe both $\xi(r)$ and $\iz(r)/\sigma_L^2$ of the mocks in real
space.

Our $b_1$ is in good agreement with the results presented in figure 16 of
ref.~\cite{gilmarin/etal:2014b}, whereas our $b_2$ is smaller than theirs,
which is $\simeq 0.95$, by 1.2$\sigma$. This may be due to the difference
in the bispectrum models. While we restrict to the local bias model and
the tree-level bispectrum, ref.~\cite{gilmarin/etal:2014b} includes a
non-local tidal bias \cite{mcdonald/roy:2009,baldauf/etal:2012,sheth/chan/scoccimarro:2012}
and uses more sophisticated bispectrum modeling using the effective $F_2$
kernel \cite{gilmarin/etal:2011,gilmarin/etal:2014a}. In \refapp{test},
we show that using the effective $F_2$ kernel and the non-local tidal
bias in the model increases the value of $b_2$, but the changes are well
within the 1-$\sigma$ uncertainties. Also, the differences of the goodness
of fit for various models are negligible.

The fitting range as well as the shapes of the bispectrum may also affect
the results: the integrated correlation function is sensitive only to the
squeezed configurations, whereas ref.~\cite{gilmarin/etal:2014b} includes
more equilateral and collapsed triangle configurations. Understanding this
difference merits further investigations.

\subsection{Measurements in redshift space}
\label{sec:mock_z}
\begin{figure}[t]
\centering
\includegraphics[width=0.495\textwidth]{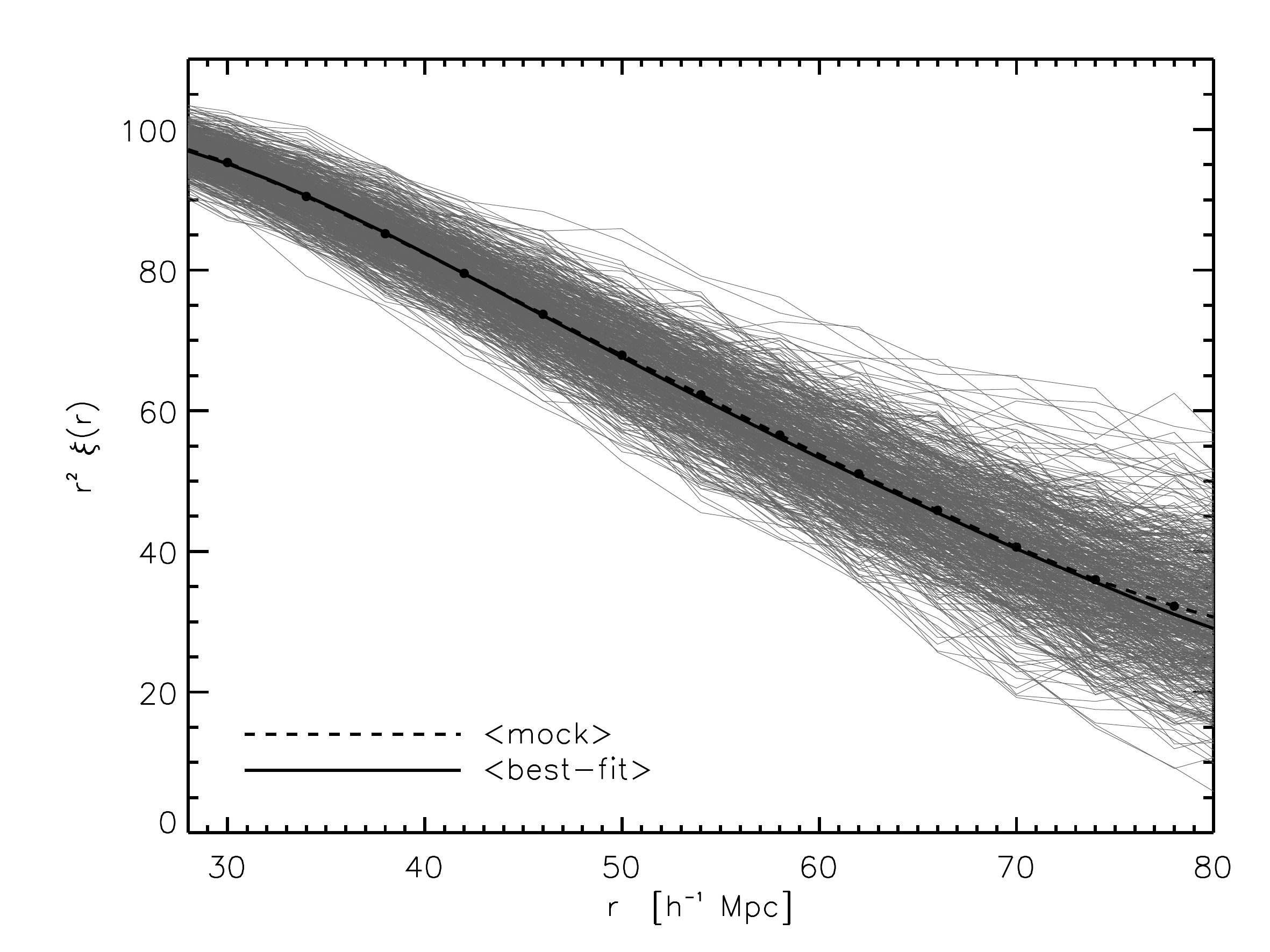}
\includegraphics[width=0.495\textwidth]{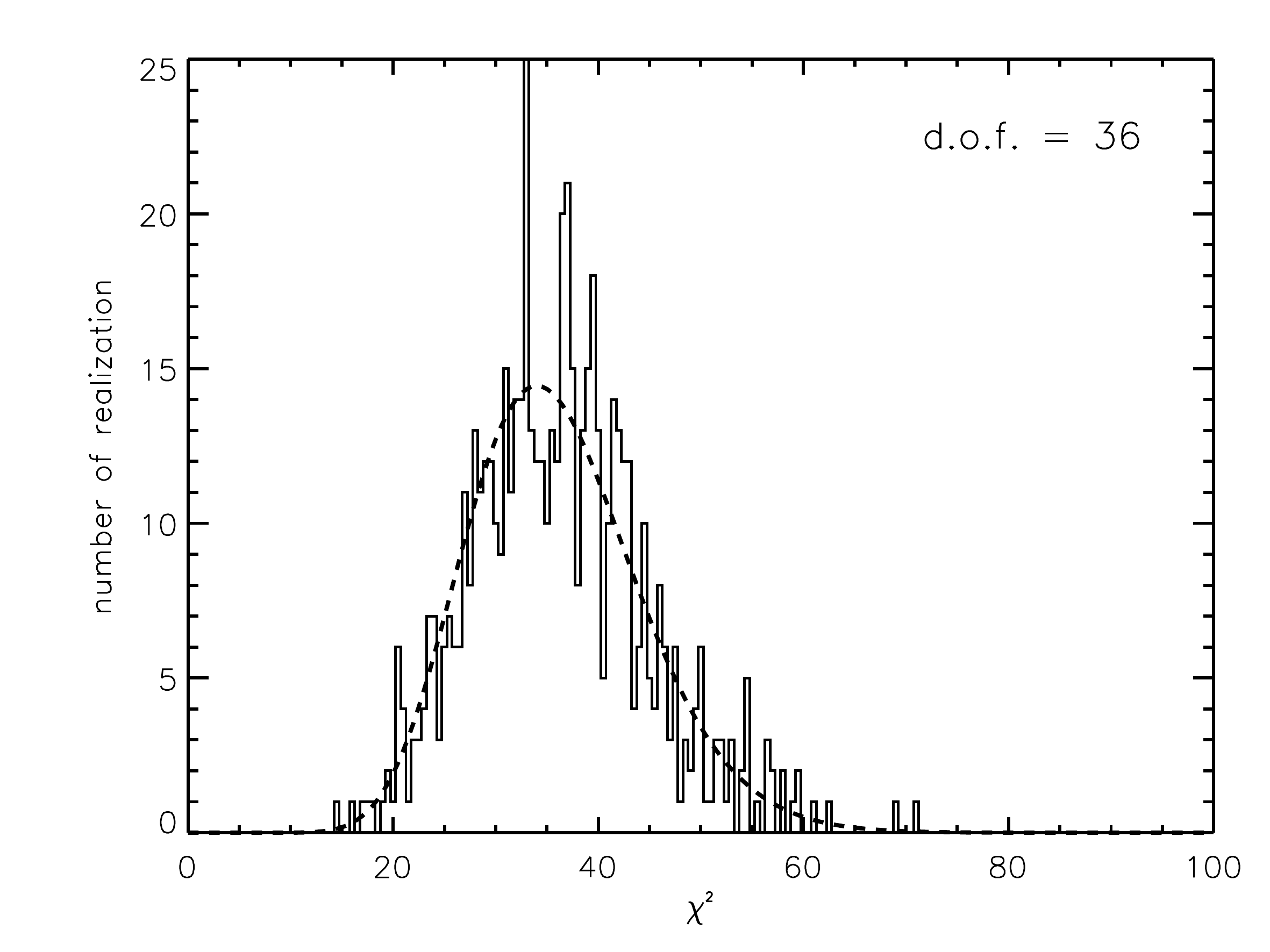}
\includegraphics[width=0.495\textwidth]{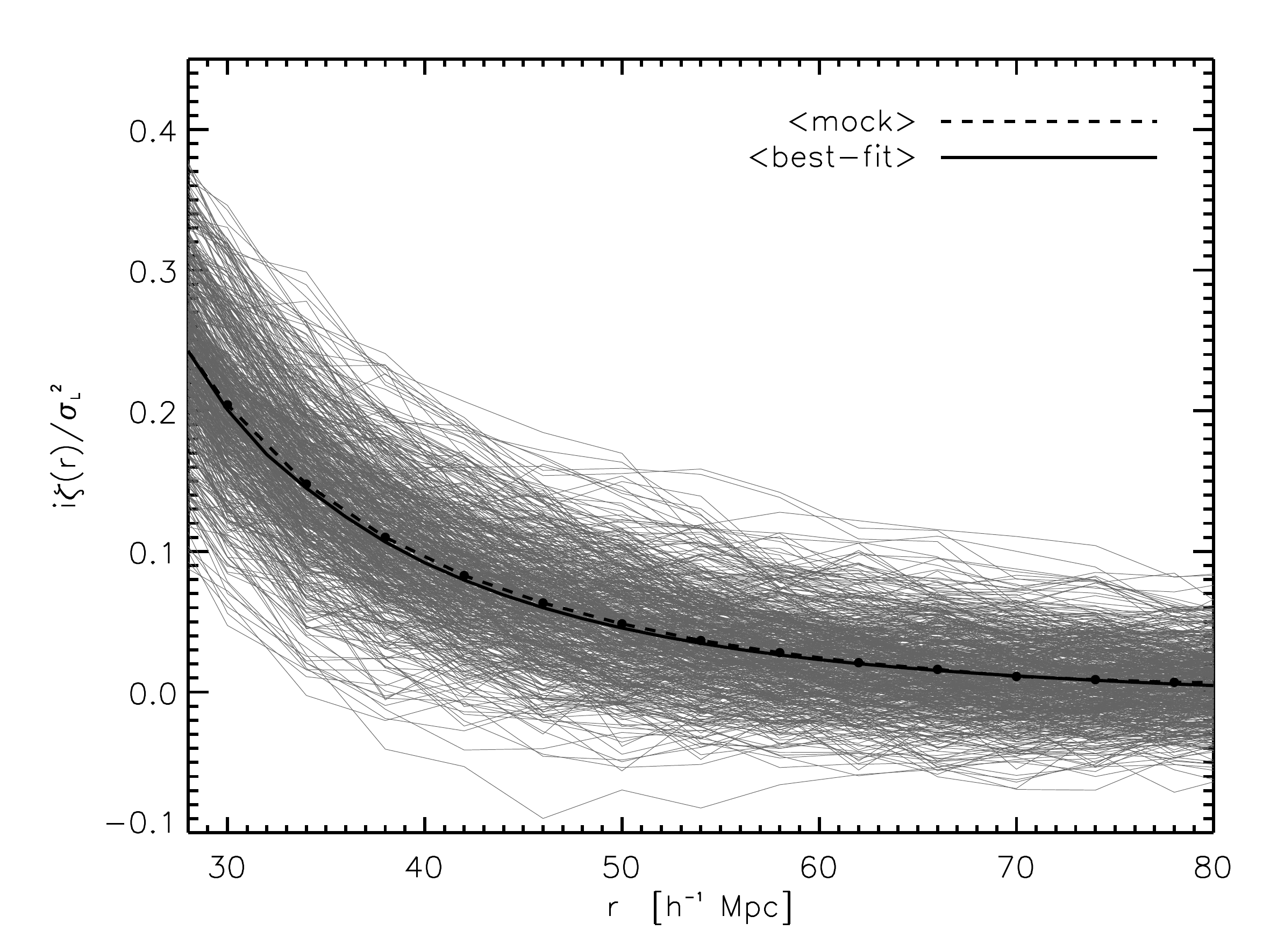}
\includegraphics[width=0.495\textwidth]{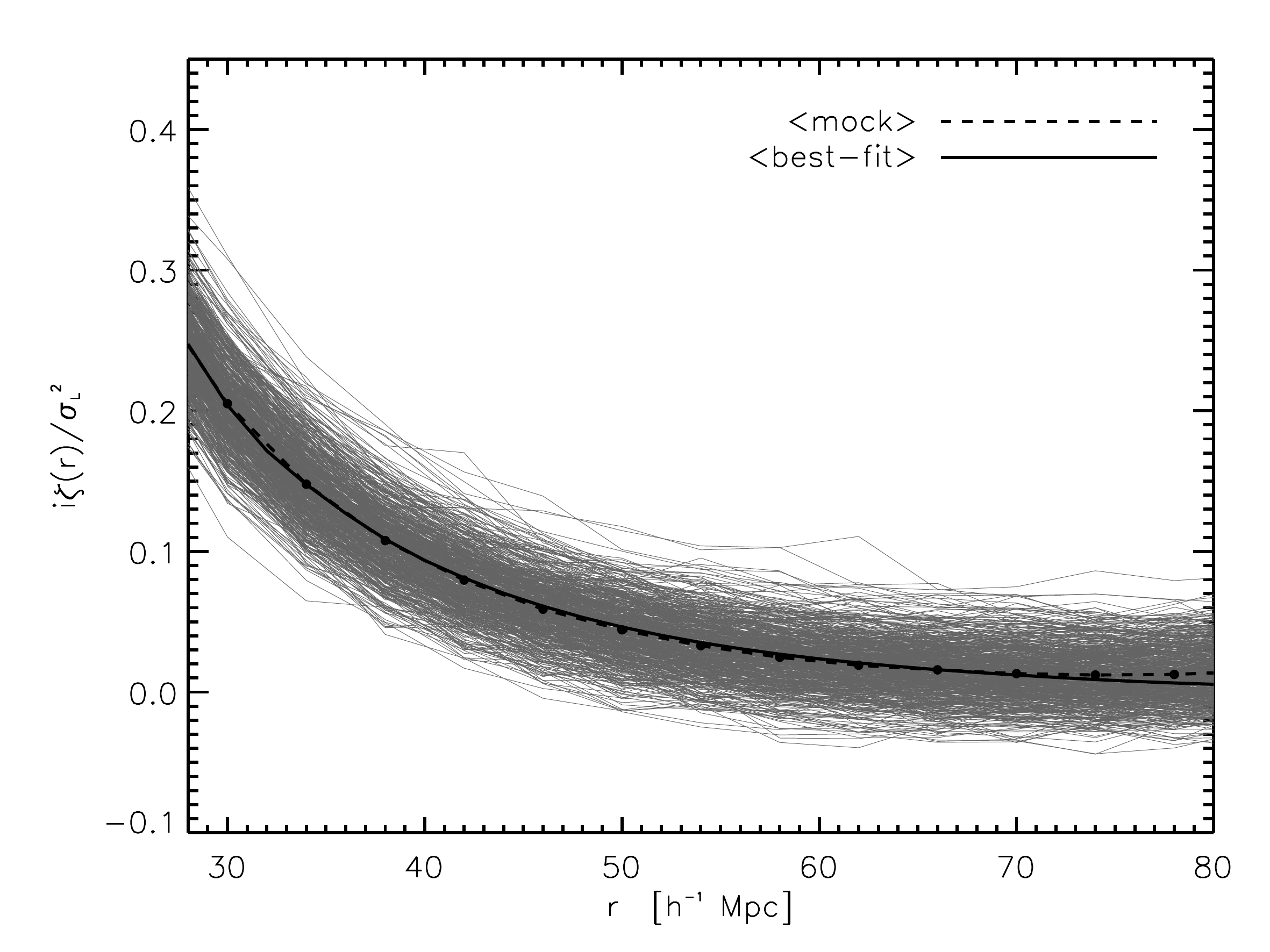}
\caption{Same as figure~\ref{fig:mock_r} but in redshift space.}
\label{fig:mock_z}
\end{figure}

\refFig{mock_z} shows the measurements of $\xi(r)$ (top left) and $\iz(r)/\sigma_L^2$
($220\hMpc$ in the bottom-left and $120\hMpc$ in the bottom-right panels) of
the mocks in redshift space. The gray lines show individual realizations, while
the  dashed lines show the mean. Similar to the analysis in \refsec{mock_r}, we
fit the models in redshift space to the measurements in $30\hMpc\le r\le78\hMpc$.
In this section, we use General Relativity to compute the growth rate,
$f(z)\approx\Omega_m(z)^{0.55}$, which yields $f(z=0.57)=0.751$. We shall allow
$f$ to vary when interpreting the measurements in the actual data.

\begin{figure}[t]
\centering
\includegraphics[width=0.415\textwidth]{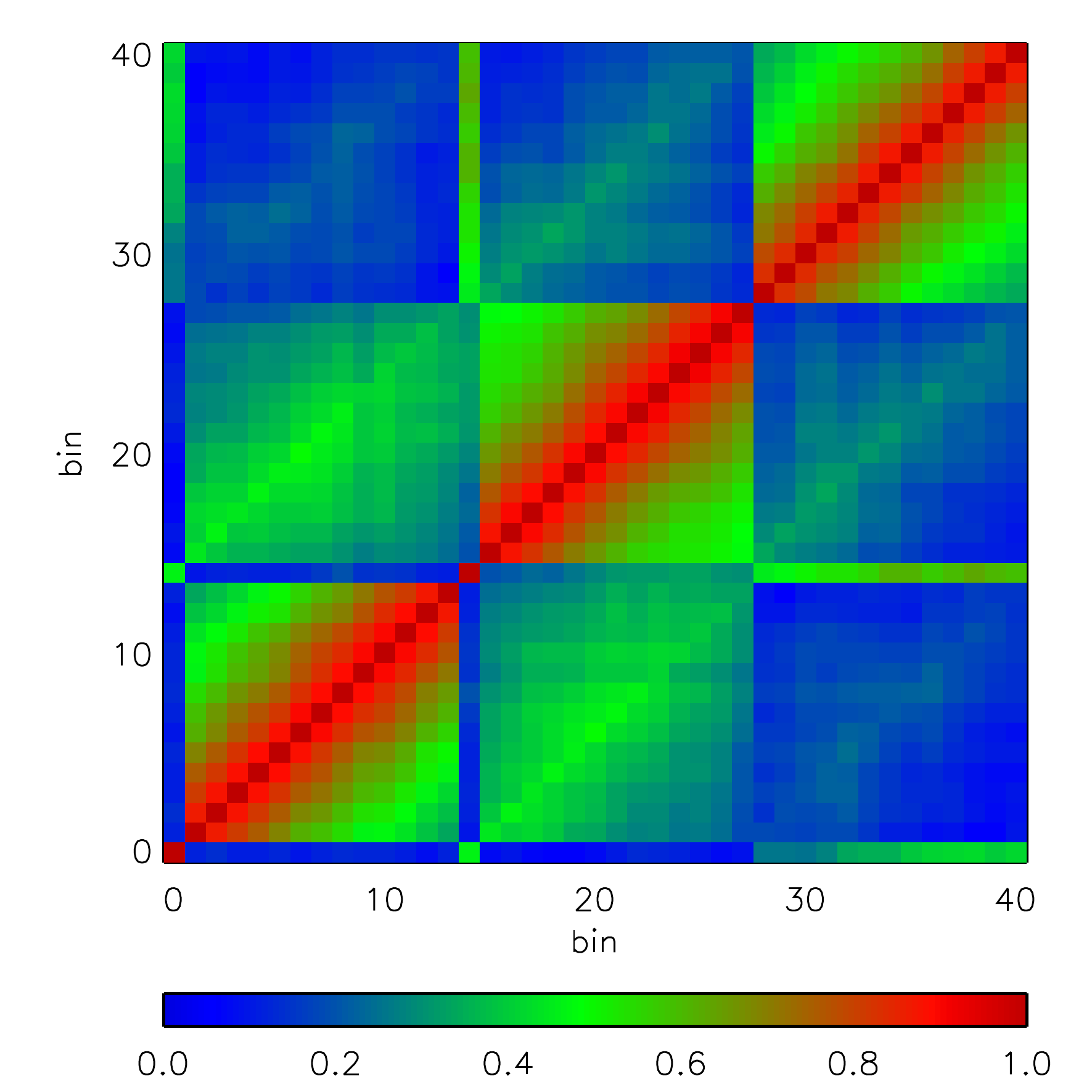}
\caption{Correlation matrix estimated from 600 mocks in redshift space. The
figure shows $\sigma_L^2$ and $\iz(r)/\sigma_L^2$ of $220\hMpc$ subvolumes
from bin 0 to 13, $\sigma_L^2$ and $\iz(r)/\sigma_L^2$ of $120\hMpc$ subvolumes
from bin 14 to 27, and $\xi(r)$ from bin 28 to 40.}
\label{fig:covred}
\end{figure}

Since there is no baryonic acoustic oscillation feature on the scales we are
interested in, we model the redshift-space two-point correlation function as
\be
 \xi_{g,z}(r)=b_1^2\left[\xi_{l,\sigma_v}(r)+A_{\rm MC}\xi_{\rm MC}(r)\right]K_p ~,
\label{eq:xiz_model}
\ee
where $\xi_{l,\sigma_v}(r)$ and $\xi_{\rm MC}(r)$ are given in \refeq{xi_model_2} and
\be
 K_p\equiv 1+\frac23\beta+\frac15\beta^2 ~,
\label{eq:kaiser}
\ee
is the Kaiser factor with $\beta\equiv f/b_1$ \cite{kaiser:1987}.
As we do not include the subdominant term proportional to $b_2$ in the two-point
function, it only gives a constraint on $b_1$, which we can then use to break the
degeneracy with $b_2$ in the integrated three-point function. We find that this
simple modeling yields unbiased $b_1$ and fulfills the demand.
We calculate the redshift-space normalized integrated three-point function using
SPT at the tree level, as described in \refsec{rsd},
and then correct for the boundary effect.
The $\sigma_L^2$ of the mocks in redshift space agrees with $b_1^2 K\,\sigma_{L,l}^2$
to percent level. The redshift-space models thus contain, as before in real space,
the three fitting parameters, $b_1$, $b_2$, and $A_{\rm MC}$. We then simultaneously
fit $\xi(r)$ and $\iz(r)/\sigma_L^2$ of both subvolumes by minimizing \refeq{chi2}.
\refFig{covred} shows the correlation matrix ($C_{ij}$ in $\chi^2$, normalized by
$\sqrt{C_{ii}C_{jj}}$) estimated from the 600 mocks in redshift space. Because we
normalize the integrated three-point function by $\sigma_L^2$, the covariance between
$\iz(r)/\sigma_L^2$ and $\sigma_L^2$ is negligible. On the other hand, the covariances
between $\iz(r)/\sigma_L^2$ and $\xi(r)$, between $\xi(r)$, and between $\iz(r)/\sigma_L^2$
for two sizes of subvolumes are significant.

The models computed with the mean of the best-fitting parameters of 600
mocks are shown as the thick solid lines in \reffig{mock_z}.
The best-fitting parameters are $b_1=1.931\pm0.077$, $b_2=0.54\pm0.35$,
and $A_{\rm MC}=1.37\pm0.82$. The agreement between the models and the
measurements in redshift space is as good as in real space.

Again, our $b_1$ is in good agreement with the results presented in figure 16 of
ref.~\cite{gilmarin/etal:2014b}, whereas our $b_2$ is smaller than theirs, which is
$\simeq 0.75$, but still well within the 1-$\sigma$ uncertainty. As noted in \refsec{mock_r},
the adopted models of the bispectrum are different. In \refapp{test}, we show that
using the effective $F_2$ and $G_2$ kernels and the non-local tidal bias in the model
increases the value of $b_2$. However, the changes are within the uncertainties,
and the goodness of the fit is similar for different models. Thus, in this paper
we shall primarily use the SPT at the tree level with local bias for simpler
interpretation of the three-point function, but also report the results for the
extended models.

%%%%%%%%%%%%%%%%%%%%%%%%%%%%%%%%%%%%%%%%%%%%%%%%%%%%%%%%%%%%%%%%%%%%%%%%%%%%

\section{Measurements of the BOSS DR10 CMASS sample}
\label{sec:data}
\begin{figure}[t]
\centering
\includegraphics[width=0.495\textwidth]{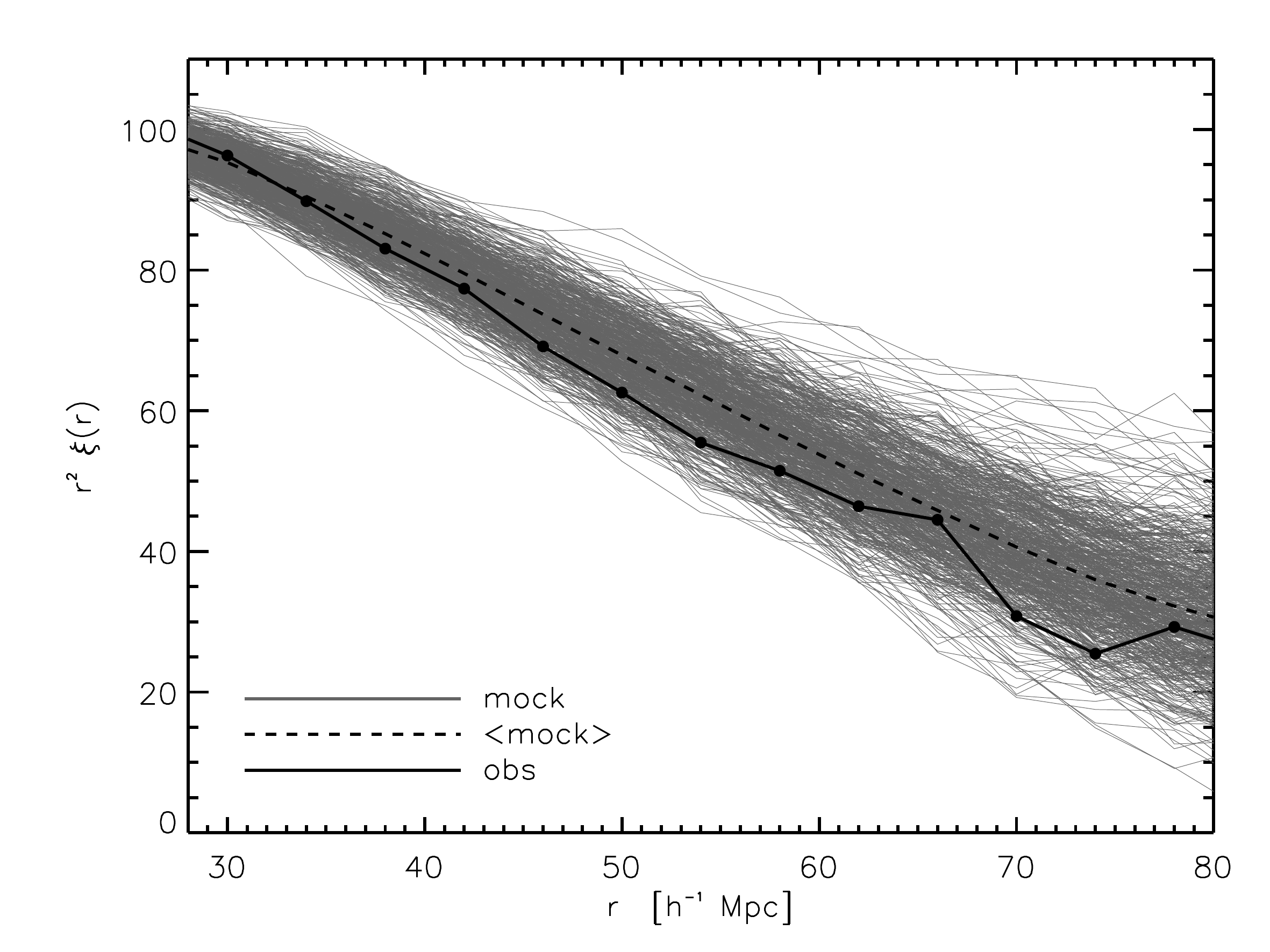}
\includegraphics[width=0.495\textwidth]{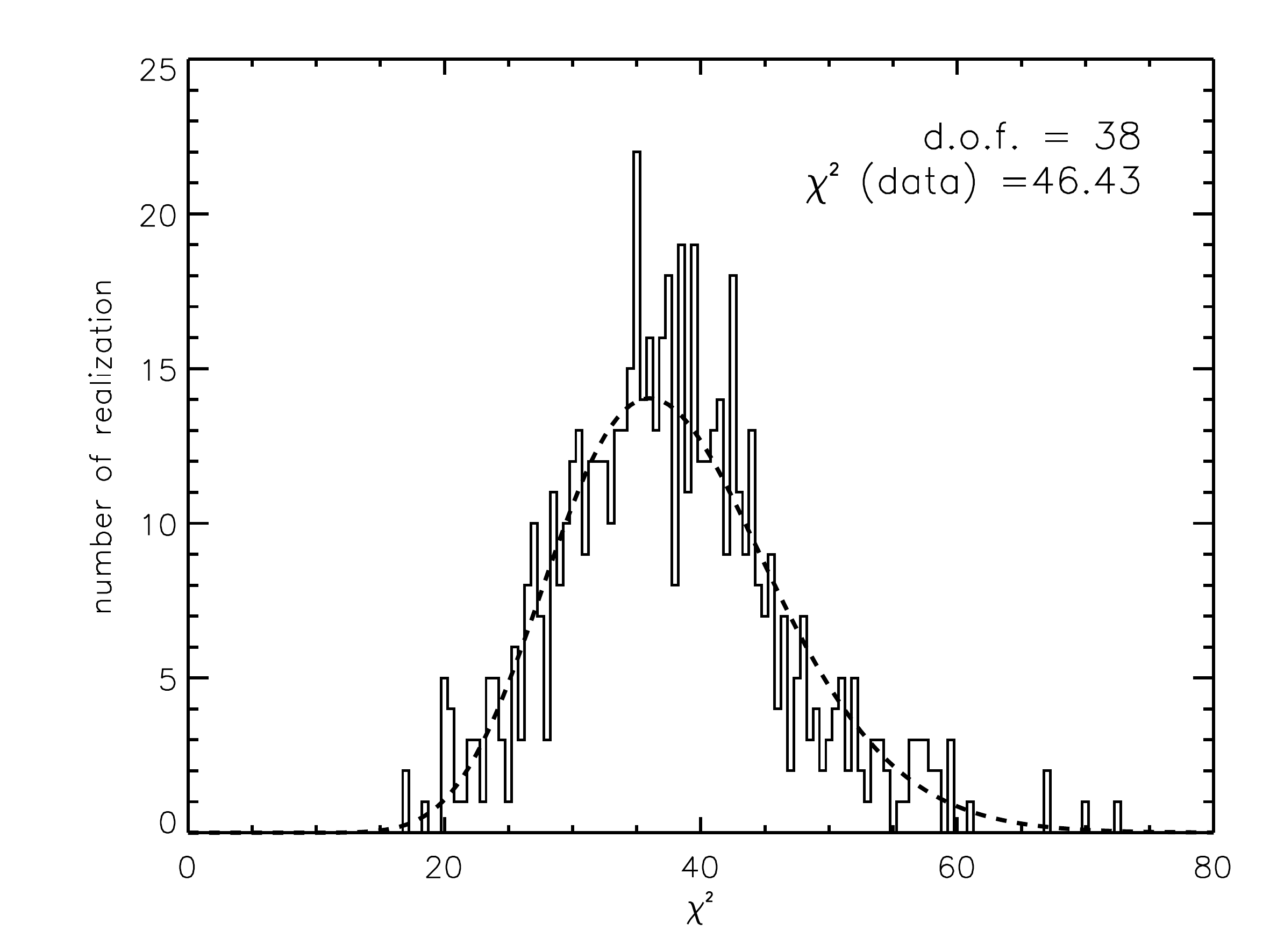}
\includegraphics[width=0.495\textwidth]{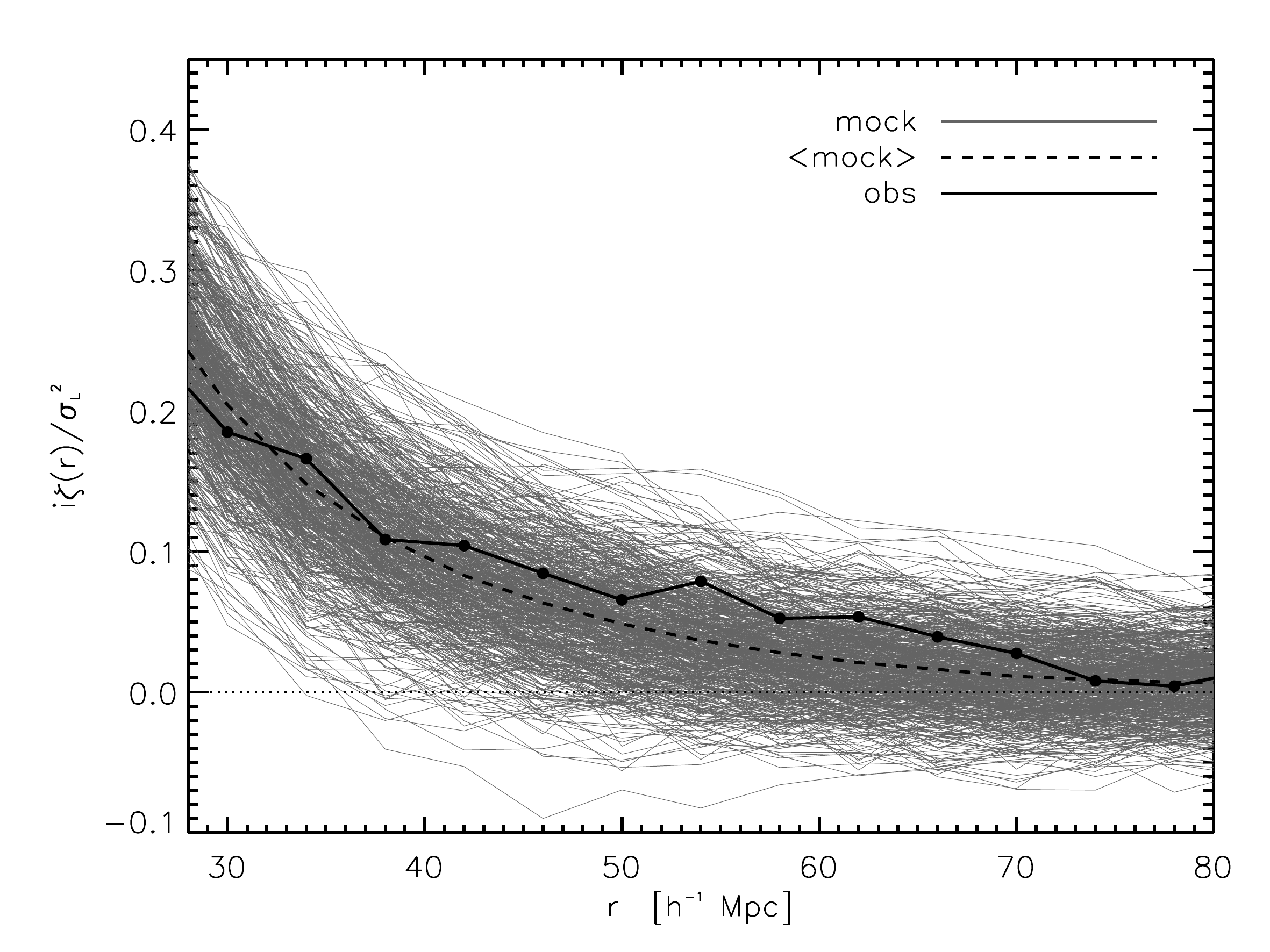}
\includegraphics[width=0.495\textwidth]{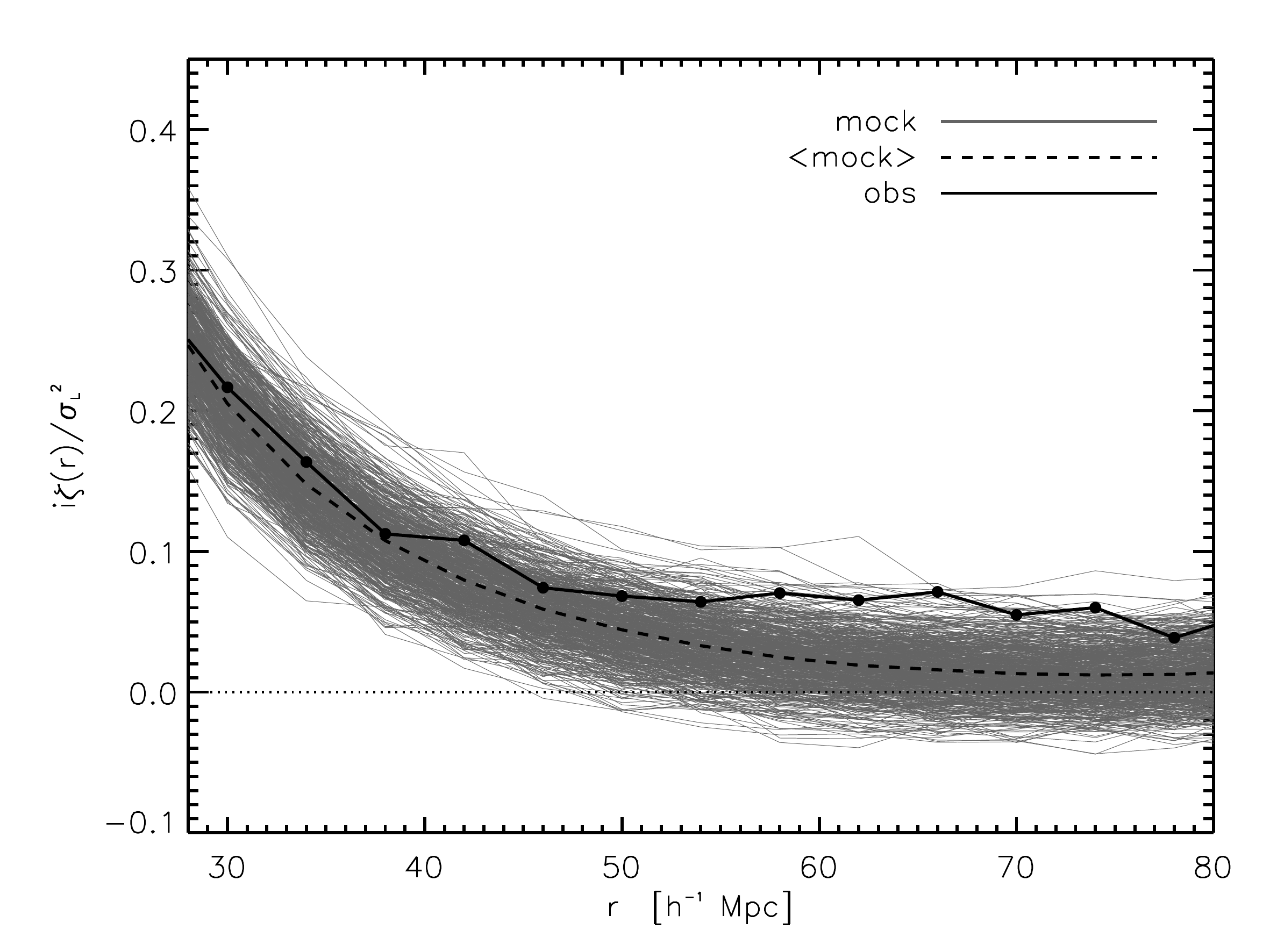}
\caption{Measurements of the BOSS DR10 CMASS sample (black solid lines).
The gray lines show individual mocks in  redshift space and the dashed line
shows the mean of mocks. (Top left) $\xi(r)$, (Bottom left) $\iz(r)/\sigma_L^2$
for $220\hMpc$ subvolumes, and (Bottom right) $\iz(r)/\sigma_L^2$ for
$120\hMpc$ subvolumes. (Top right) $\chi^2$-histogram of the 600 mocks
jointly fitting the three amplitudes to $\xi(r)$ and $\iz(r)/\sigma_L^2$
in redshift space. The dashed line shows the $\chi^2$-distribution with
d.o.f.=38. The $\chi^2$ value measured from the BOSS DR10 CMASS sample is 46.4.}
\label{fig:data}
\end{figure}

We now present measurements of the position-dependent correlation function from
the BOSS DR10 CMASS sample\footnote{Catalogs of galaxies and the random samples
can be found in \url{http://www.sdss3.org}.} in NGC. The detailed description
of the observations can be found in refs.~\cite{ahn/etal:2013,anderson/etal:2013}.
Briefly, the sample contains 392,372 galaxies over 4,892 deg$^2$ in the redshift
range of $0.43<z<0.7$, which corresponds to the comoving volume of approximately
$2~h^{-3}~{\rm Gpc}^3$. We also weight the galaxies by $w_{\rm BOSS}$ to correct
for the observational systematics. We follow \refsec{division} to divide the
observations into subvolumes. However, the observations have their own set of
random samples, which are different from the ones of the mocks (the random samples
of the mocks have slightly higher $\bar n$ and different $\bar n(z)$), so we adjust
the redshift cuts to be $z_{\rm cut}=0.5108$, 0.5717 and $z_{\rm cut}=0.48710$,
0.52235, 0.55825, 0.60435 for the two resolutions, respectively. The resulting
properties of subvolumes of the observations and mocks are similar.

The mocks are constructed to match the two-point function of the observed galaxies,
but not for the three-point function. Hence there is no guarantee that the three-point
function of mocks agrees with the observations. We can test this using our measurements.  

\begin{table}[t]
\centering
\begin{tabular}{ | c | c c c | }
\hline
& avg$[\sigma_{L,\rm mock}^2]$ & var$[\sigma_{L,\rm mock}^2]$ & $\sigma_{L,\rm data}^2$ \\
\hline
 $220\hMpc$ & $4.6\times10^{-3}$ & $5.6\times10^{-4}$ & $4.9\times10^{-3}$ \\
 $120\hMpc$ & $2.4\times10^{-2}$ & $1.3\times10^{-3}$ & $2.5\times10^{-2}$ \\
\hline
\end{tabular}
\caption{Measurements of $\sigma_L^2$ of the mock catalogs and the BOSS DR10 CMASS sample.}
\label{tab:sigma2}
\end{table}

The measurements of $\xi(r)$ and $\iz(r)/\sigma_L^2$ from the observations are
shown as the solid lines in \reffig{data}; the measurements of $\sigma_L^2$ is
summarized in \reftab{sigma2}. The measurements are consistent visually
with the mocks within the scatter of the mocks\footnote{These measurements of
$\iz(r)/\sigma_L^2$ are done for one effective redshift. We compare $\iz(r)/\sigma_L^2$
of the observations and mocks in different redshift bins in \refapp{iz_zevolve},
finding that the observations and mocks are consistent at all redshift bins to
within the scatter of the mocks.}, and we shall quantify the goodness of fit
using $\chi^2$ statistics later.

To quantify statistical significance of the detection of $\iz(r)/\sigma_L^2$
and the goodness of fit, we use the mean of the mocks as the model (instead
of the model based on perturbation theory used in section~\ref{sec:mock_z}),
and fit only the amplitudes of $\iz(r)/\sigma_L^2$, $\xi(r)$, and $\sigma_L^2$
to the observations and the 600 mocks by minimizing \refeq{chi2}. Specifically,
we use $O_i(r) = A_i\,O_i^{\rm mock}(r)$ as the model,  where $O_1(r) = \iz(r)/\sigma_L^2$,
$O_2(r)=\xi(r)$, and $O_3 = \sigma_L^2$, with the amplitudes $A_1,\,A_2,\,A_3$. 

\begin{table}[t]
\centering
\begin{tabular}{ | c | c c c | }
\hline
& $A_1$ & $A_2$ & $A_3$ \\
\hline
1-$\sigma$ error & 0.12 & 0.03 & 0.04 \\
best-fit (DR10) & 0.89 & 1.02 & 1.08 \\
\hline
\end{tabular}
~~~~~~~
\begin{tabular}{ | c | c c c | }
\hline
& $(A_1,A_2)$ & $(A_1,A_3)$ & $(A_2,A_3)$ \\
\hline
corr & 0.34 & 0.09 & 0.36 \\
\hline
\end{tabular}
\caption{Results of fitting the amplitudes: $A_1$ is $\iz(r)/\sigma_L^2$, $A_2$
is $\xi(r)$, and $A_3$ is $\sigma_L^2$. (Left) The 1-$\sigma$ uncertainties of
the amplitudes estimated from the mocks, and the best-fitting amplitudes of BOSS
DR10 CMASS sample with respect to the mean of the mocks. (Right) The correlation
coefficients of the amplitudes.}
\label{tab:fit_amp}
\end{table}

\refTab{fit_amp} summarizes the fitted amplitudes. The 1-$\sigma$ uncertainties
and the correlations are estimated from the 600 mocks. Since we normalize $\iz(r)$
by $\sigma_L^2$, the correlation between $A_1$ and $A_3$ is small. On the other hand,
$A_2$ and $A_3$ are correlated significantly because $\sigma_L^2$ is an integral of
the two-point function [\refeq{sigmalL2}]. 

Comparing the BOSS DR10 CMASS sample to the mean of the mocks, we find that
$\iz(r)/\sigma_L^2$ is 1-$\sigma$ lower, 
$\xi(r)$ is unbiased (by construction of the mocks), and $\sigma_L^2$ is 2-$\sigma$
higher. The result of $A_1$ for the data is driven by the correlation between
different separations of $\iz(r)/\sigma_L^2$.
On the other hand, the result of $A_3$ is driven by the positive correlation
between $\xi(r)$ and $\sigma_L^2$.
While $\sigma_L^2$ of the data for two subvolumes are larger than that of the mocks
but still at the boundary of the variances (see \reftab{sigma2}), it requires an even
higher $A_3$ to minimize $\chi^2$ when we jointly fit the three amplitudes.
The fact that $A_3$ is larger than $A_2$ is also possibly due
the contributions to $\sigma_L^2$ from small separations (including
stochasticity at zero separations), where the mocks were not optimized.
We find $A_1=0.89\pm 0.12$, i.e., a 7.4$\sigma$ detection of the
integrated three-point function of the BOSS DR10 CMASS sample. 

In order to assess the goodness of fit,  we use the distribution of $\chi^2$,
a histogram of which is shown in the top right panel of \reffig{data}. In total
there are 41 fitting points (13 fitting points for $\xi(r)$ and two sizes of
subvolumes for $\iz(r)/\sigma_L^2$, and two fitting points for $\sigma_L^2$)
with three fitting parameters, so d.o.f.=38. The $\chi^2$ value of the observations
is 46.4, and the probability to exceed this $\chi^2$ value is more than 16\%.
Given the fact that the mocks are constructed to match only the two-point function
of the observations, this level of agreement for both the two-point and integrated
three-point correlation functions is satisfactory.

%%%%%%%%%%%%%%%%%%%%%%%%%%%%%%%%%%%%%%%%%%%%%%%%%%%%%%%%%%%%%%%%%%%%%%%%%%%%

\section{Cosmological interpretation of the integrated three-point function}
\label{sec:interpretation}

What can we learn from the measured $\iz(r)/\sigma_L^2$? In section~\ref{sec:mock_z},
we show that the prediction for $\iz(r)/\sigma_L^2$ based on SPT at the tree-level
in redshift space provides an adequate fit to the mocks to within the scatter of
the mocks; thus, we can use this prediction to infer cosmology from $\iz(r)/\sigma_L^2$.
Note that any unmodeled effects in the integrated three-point function such as
nonlinearities of the matter density, nonlocal bias parameters, and redshift-space
distortions beyond the Kaiser factor, will tend to bias our measurement of cosmological
parameters based on $\iz(r)$. We will discuss caveats at the end of this section.

Since the linear two-point and the tree-level three-point functions are proportional
to $\se^2$ and $\se^4$, respectively, and $\sigma_L^2$ is proportional to $\se^2$,
the scaling of the redshift-space correlation functions is
\ba
 \xi_{g,z}(r)\:&=b_1^2K\left[\xi^{\rm fid}_{l,\sigma_v}(r)\left(\frac{\se}{\sef}\right)^2
 +A_{\rm MC}\xi^{\rm fid}_{\rm MC}(r)\left(\frac{\se}{\sef}\right)^4\right]\,, \vs
 \frac{\iz_{g,z}(r)}{\sigma_L^2}\:&=
\frac{\iz^{\rm fid}_{g,z}(r)}{b_1^2\sigma_{L,l}^2K_p}\left(\frac{\se}{\sef}\right)^2
 \frac1{f_{\rm bndry}(r)} ~,
\label{eq:obs_zspace}
\ea
where ``fid'' denotes the quantities computed with the fiducial value of $\sigma_8$.
Note that $\xi_{\rm MC}(r)$ is proportional to $\se^4$ because it is an integral
of two linear power spectra (see \refeq{xi_model}). Since $\xi_{l,\sigma_v}(r)$
dominates the signal, the parameter combinations $b_1\se$ and $K=1+2\beta/3+\beta^2/5$
are degenerate in the two-point  function. That is, the amplitude of the two-point
function measures only $(b_1\se)^2+\frac{2}{3}(b_1\se)(f\se)+\frac{1}{5}(f\se)^2$.
This degeneracy can be lifted by including the quadrupole of the two-point function
in redshift space. See refs.~\cite{samushia/etal:2013,tojeiro/etal:2014,sanchez/etal:2013,beutler/etal:2013}
for the latest measurements using the BOSS DR11 sample.

As for the three-point function, \reffig{iz_norm} shows that the $b_1^3$ and
$b_1^2 b_2$ terms are comparable for $b_1\approx b_2$. This means that, at the
three-point function level, the nonlinear bias appears in the leading order,
so the amplitude of the three-point function measures a linear combination of
$b_1$ and $b_2$. This provides a wonderful opportunity to determine $b_2$. The
challenge is to break the degeneracy between $b_2$, $b_1$, $f$, and $\sigma_8$.
For this purpose, we combine our results with the two-point function in redshift
space and the weak lensing measurements of BOSS galaxies. We take the constraints
on $b_1\se(z=0.57)=1.29\pm0.03$ and $f(z=0.57)\se(z=0.57)=0.441\pm0.043$ from table~2
in ref.~\cite{samushia/etal:2013}. To further break the degeneracy between $b_1$,
$f$, and $\se$, we take the constraint on  $\se=0.785\pm0.044$ from ref.~\cite{miyatake/etal:2013,more/etal:2014},
where they jointly analyze the clustering and the galaxy-galaxy lensing using the
BOSS DR11 CMASS sample and the shape catalog from Canada France Hawaii Telescope
Legacy Survey.

We assume Gaussian priors on $b_1\se$, $f\se$, and $\se$ with the known covariance
between $b_1\se$ and $f\se$. The cross-correlation coefficient between $b_1\se$
and $f\se$ is $-0.59$, as shown in figure~6 of ref.~\cite{samushia/etal:2013}.
We then run the Markov Chain Monte Carlo with the Metropolis-Hastings algorithm
to fit the model \refeq{obs_zspace} to the observed $\iz(r)/\sigma_L^2$.
We find $b_2=0.41\pm0.41$, and the results for the extended models are summarized
in \reftab{b2_data}.

\begin{table}[t]
\centering
\begin{tabular}{ | c | c c c c | }
\hline
 & baseline & eff kernel & tidal bias & both \\
\hline
 $b_2$ & $0.41\pm0.41$ & $0.51\pm0.41$ & $0.48\pm0.41$ & $0.60\pm0.41$ \\
\hline
\end{tabular}
\caption{Best-fitting $b_2$ and their uncertainties for BOSS DR10 CMASS sample
for the extended models. The detailed description of the extended models is
in \refapp{test}.}
\label{tab:b2_data}
\end{table}

The value of $b_2$ we find is lower than the mean of the mocks, $b_2^{\rm mock}=0.54\pm0.35$.
The difference is mainly due to two reasons. First, the amplitude of the integrated
three-point function of the observations is lower than that of the mocks by 10\%
($A_1=0.89\pm0.12$).
Second, the priors from the correlation function and lensing constraint $b_1$ to
be close to 2.18, which is larger than that of the mocks, $b_1^{\rm mock}=1.93$.  
Thus, it requires a smaller $b_2$ to fit the three-point function.
The argument is similar for the extended models. Note, however, that the nonlinear
bias of the data is still statistically consistent with the mocks.

Let us conclude this section by listing three caveats regarding our cosmological
interpretation of the measured integrated three-point function.
\begin{enumerate}
\item The models we use, \refeq{obs_zspace}, are based on tree-level perturbation
		theory, the lowest order redshift-space distortion treatment, as well as
		on the local bias parametrization. While this simple model describes the
		mocks well, as shown in \refsec{mock_r} and \ref{sec:mock_z}, we discuss
		in \refapp{test} that using the effective $F_2$ and $G_2$ and the non-local
		tidal bias brings $b_2$ closer to that of ref.~\cite{gilmarin/etal:2014b}.
		We, however, find similar goodness of fit for various models, and thus
		we cannot distinguish between these models.

\item Covariances between the integrated three-point function, monopole and quadrupole
		two-point function, and weak lensing signals are ignored in our treatment.
		This can and should be improved by performing a joint fit to all the observables.

\item The cosmology is fixed throughout the analysis, except for $f$ and $\sigma_8$.
        In principle, marginalizing over the cosmological parameters is necessary
        to obtain self-consistent results, although the normalized integrated three-point
        function is not sensitive to cosmological parameters such as $\Omega_m$ as
        shown in figure~6 of ref.~\cite{chiang/etal:2014}.

\end{enumerate}
These caveats need to be addressed in the future work.

%%%%%%%%%%%%%%%%%%%%%%%%%%%%%%%%%%%%%%%%%%%%%%%%%%%%%%%%%%%%%%%%%%%%%%%%%%%%

\section{Conclusions}
\label{sec:conclusion}
In this paper, we have reported on the first measurement of the three-point
function with the position-dependent correlation function from the SDSS-III
BOSS DR10 CMASS sample. The correlation between the position-dependent correlation
function measured within subvolumes and the mean overdensities of those subvolumes
is robustly detected at 7.4$\sigma$. This correlation measures the integrated
three-point function, which is the Fourier transform of the integrated bispectrum
introduced in ref.~\cite{chiang/etal:2014}, and is sensitive to the bispectrum
in the squeezed configurations.

Both the position-dependent correlation function and the mean overdensity are easier
to measure than the three-point function.  The computational expense for the two-point
function is much cheaper than the three-point function estimator using the triplet-counting
method. In addition, for a fixed size of the subvolume, the integrated three-point
function depends only on one variable (i.e., separation), unlike the full three-point
function which depends on three separations. This property allows for a useful compression
of information in the three-point function in the squeezed configurations, and makes
physical sense because the integrated three-point function measures how the small-scale
two-point function, which depends only on the separation, responds to a long-wavelength
fluctuation \cite{chiang/etal:2014}. As there are only a small number of  measurement
bins, the covariance matrix of the integrated three-point function is easier to estimate
than that of the full three-point function from a realistic number of mocks. We have
demonstrated this advantage in the paper.

Of course, since this technique measures the three-point function with one long-wavelength
mode (mean overdensity in the subvolumes) and two relatively small-wavelength modes
(position-dependent correlation function), it is not sensitive to the three-point
function of other configurations, which were explored by ref.~\cite{gilmarin/etal:2014b}.

We have used the mock galaxy catalogs, which are constructed to match the two-point
function of the SDSS-III BOSS DR10 CMASS sample in redshift space, to validate our
method and theoretical model. We show that in both real and redshift space, the
integrated three-point function of the mocks can be well described by the tree-level
SPT model. However, the nonlinear bias which we obtain from the mocks is higher than
that reported in ref.~\cite{gilmarin/etal:2014b}. This is possibly due to the differences
in the scales and configurations of the three-point function used for the analyses.
As discussed in \refsec{interpretation}, any unmodeled nonlinear effects in the redshift-space
integrated three-point function of CMASS galaxies will tend to bias $b_2$, and will
bias this parameter differently  than the measurement of ref.~\cite{gilmarin/etal:2014b}.

Taking the mean of the mocks as the model, and treating the amplitudes of two- and
three-point functions as free parameters, we find  the best-fit amplitudes of
$\iz(r)/\sigma_L^2$, $\xi(r)$, and $\sigma_L^2$ of the CMASS sample. With respect to
the mean of the mocks, the observations show a somewhat smaller $\iz(r)/\sigma_L^2$
($A_1=0.89\pm 0.12$) and larger $\sigma_L^2$, while the ensemble two-point function
$\xi(r)$ matches the mocks. Given that the mocks are generated to match specifically
the two-point function of the BOSS DR10 CMASS sample within a certain range of
separations, the level of agreement between the observations and mocks is satisfactory.

Finally, by combining the integrated three-point function and the constraints
from the anisotropic clustering  ($b_1\se$ and $f\se$ in \cite{samushia/etal:2013})
and from the weak lensing measurements ($\se$ in \cite{more/etal:2014}), we break
the degeneracy between $b_1$, $b_2$, $f$, and $\sigma_8$. We find $b_2=0.41\pm0.41$
for the BOSS DR10 CMASS sample. The caveat of this result is that our model,
\refeq{obs_zspace}, relies on a rather simple model in redshift space as well
as on the local bias parametrization. We leave the extension of the model to
improved bias and redshift-space distortion modeling (especially in light of the
comparison with the results in ref.~\cite{gilmarin/etal:2014b}) for future work.

In summary, we have demonstrated that the integrated three-point function is
a new observable which can be measured straightforwardly from galaxy surveys
using basically the existing and routinely applied machinery to compute the
two-point function, and has the potential to yield a useful constraint on the
quadratic nonlinear bias parameter. Moreover, since the integrated three-point
function is most sensitive to the bispectrum in the squeezed configurations,
it is sensitive to primordial non-Gaussianity of the local type (parametrized
by $f_{\rm NL}$), thereby offering a probe of the physics of inflation. We plan
to extend this work to search for the signature of primordial non-Gaussianity
in the full BOSS galaxy sample.

\acknowledgments
We would like to thank Marc Manera for sharing the mock catalogs, and Lado Samushia
for computing the correlation between $b_1\sigma_8$ and $f\sigma_8$. We would also
like to thank Masahiro Takada, Shun Saito, and Surhud More for useful discussions.
We would like to thank an anonymous referee for useful comments.

Funding for SDSS-III has been provided by the Alfred P. Sloan Foundation, the
Participating Institutions, the National Science Foundation, and the U.S. Department
of Energy Office of Science. The SDSS-III web site is http://www.sdss3.org/.

SDSS-III is managed by the Astrophysical Research Consortium for the Participating
Institutions of the SDSS-III Collaboration including the University of Arizona,
the Brazilian Participation Group, Brookhaven National Laboratory, Carnegie Mellon
University, University of Florida, the French Participation Group, the German
Participation Group, Harvard University, the Instituto de Astrofisica de Canarias,
the Michigan State/Notre Dame/JINA Participation Group, Johns Hopkins University,
Lawrence Berkeley National Laboratory, Max Planck Institute for Astrophysics,
Max Planck Institute for Extraterrestrial Physics, New Mexico State University,
New York University, Ohio State University, Pennsylvania State University,
University of Portsmouth, Princeton University, the Spanish Participation Group,
University of Tokyo, University of Utah, Vanderbilt University, University of
Virginia, University of Washington, and Yale University. 

%%%%%%%%%%%%%%%%%%%%%%%%%%%%%%%%%%%%%%%%%%%%%%%%%%%%%%%%%%%%%%%%%%%%%%%%%%%%
\appendix

\section{Testing the integrated three-point function estimator
with Gaussian realizations and the local bias model}
\label{app:gaussian}

\begin{figure}[t]
\centering
\includegraphics[width=0.495\textwidth]{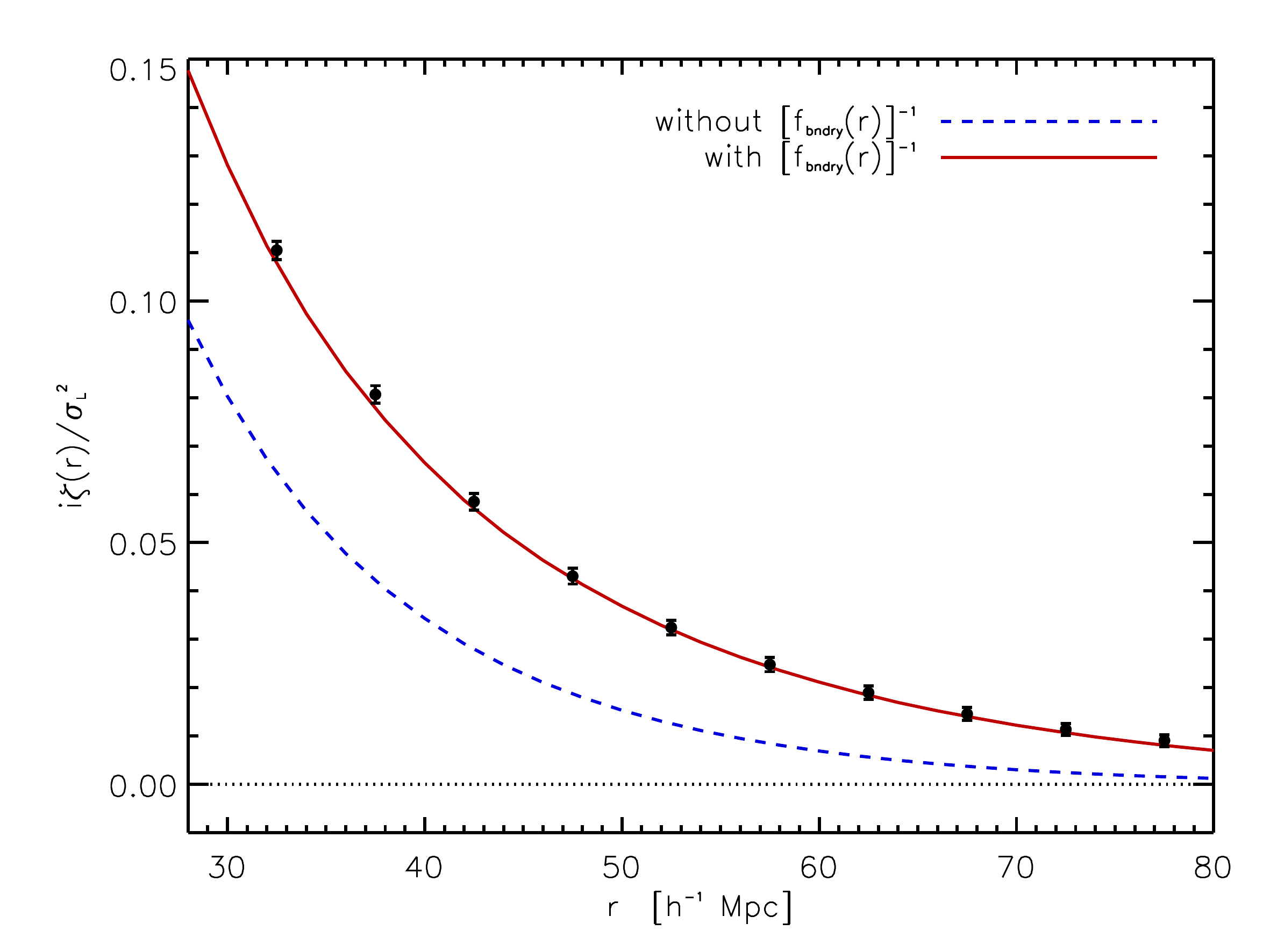}
\includegraphics[width=0.495\textwidth]{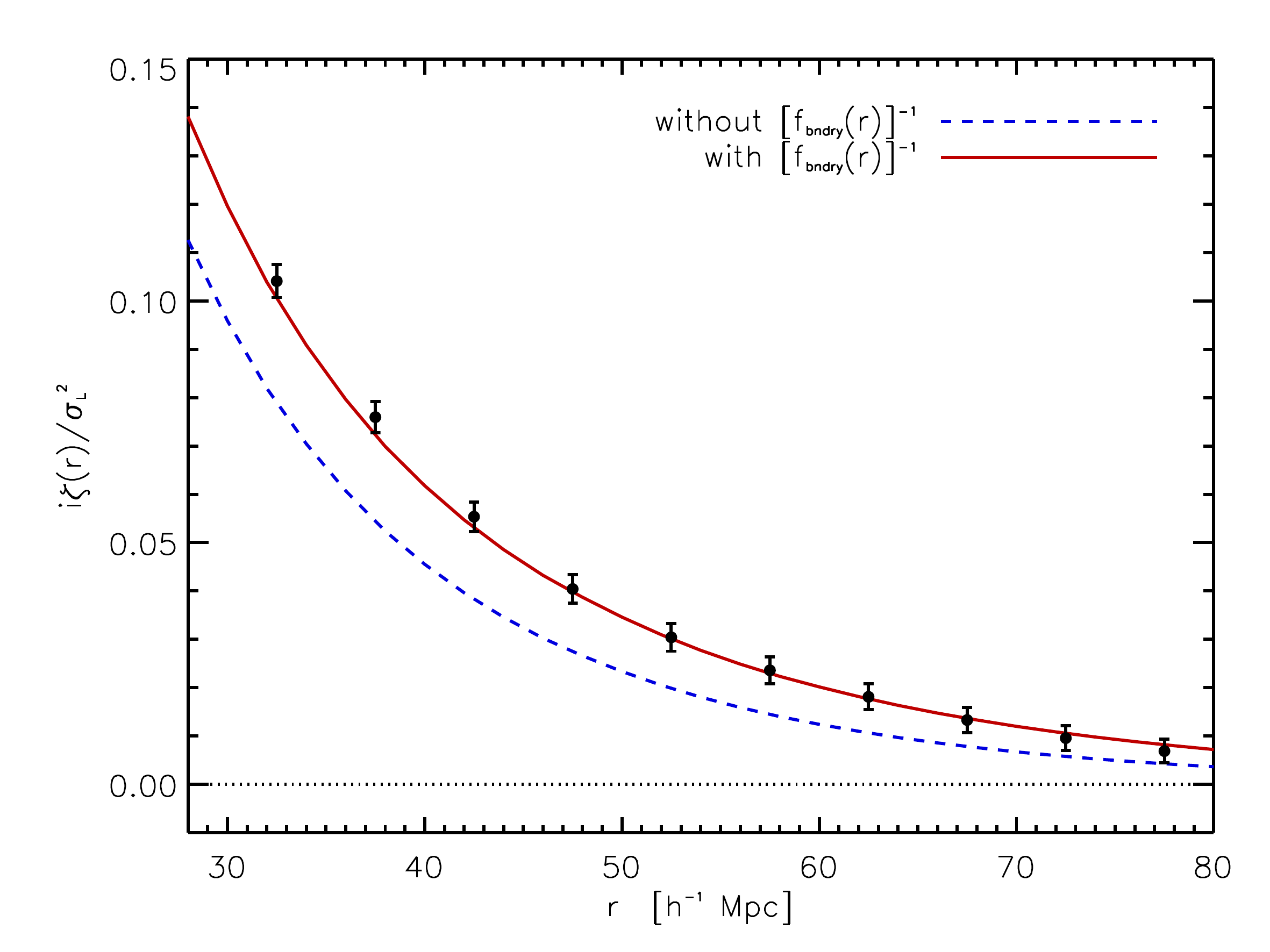}
\caption{The normalized integrated three-point functions of the mock halo density
field with $b_1=3$ and $b_2=1$. The left and right panels are for $V_L=100~h^{-3}~{\rm Mpc}^3$
and $200~h^{-3}~{\rm Mpc}^3$, respectively. The data points show the mean of 300
Gaussian realizations, and the error bars are the variances of the mean (but note
that the data points are highly correlated). The blue dashed and red solid lines
are the theoretical models ($\iz_{b_2}$) without and with $[f_{\rm bndry}(r)]^{-1}$,
respectively.}
\label{fig:mock_halo}
\end{figure}

We shall demonstrate that our integrated three-point function estimator is unbiased.
To do this, we first generate the matter density field, $\d_m(\vr)$, by Gaussian
realizations\footnote{Since $\d_m(\vr)$ follows the Gaussian statitistics, it is
possible that $\d_m(\vr)<-1$, which is unphysical. However, as we only compute the
power spectrum without Poisson sampling the density field, this effect can be neglected.}
with $P_l(k)$ at $z=0$ for the volume $V_r$ of $1200~h^{-3}~{\rm Mpc}^3$ and a mesh
size of $4\hMpc$. We then compute a mock ``halo'' density field using the local bias
model via
\be
 \d_h(\vr)=b_1\d_m(\vr)+\frac{b_2}2\left[\d_m^2(\vr)
 -\frac{\sum_{\vr\in V_r}\d_m^2(\vr)}{\sum_{\vr\in V_r}1}\right] ~,
\ee
where we set $b_1=3$ and $b_2=1$, and $\sum_{\vr\in V_r}$ denotes a sum over grid cells
in the entire volume. Note that $\sum_{\vr\in V_r}\d_h(\vr)=0$. We then divide the entire
volume $V_r$ into $N_s=12^3=1728$ subvolumes $V_L$ of $100~h^{-3}~{\rm Mpc}^3$ and $N_s=6^3=216$
subvolumes $V_L$ of $200~h^{-3}~{\rm Mpc}^3$. The two-point function in the subvolumes
and the integrated three-point function are estimated by
\be
 \xi_h(r,\vr_L)=\frac{\sum_{\vx+\vr,\vx\in V_L}\d_h(\vx+\vr)\d_h(\vx)}{\sum_{\vx+\vr,\vx\in V_L}1} ~,~~
 \iz_h(r)=\sum_{i=1}^{N_s}\xi_h(r,\vr_L)\bar\d_h(\vr_L) ~,
\label{eq:xih_izh}
\ee
where $\bar\d_h(\vr_L)$ is the mean halo overdensity in the subvolume centered at $\vr_L$.
Note that the denominator in the estimator of $\xi_h(r,\vr_L)$ takes the boundary effect
into account so $\langle\xi_h(r,\vr_L)\rangle=\xi_h(r)$ without $f_{\rm bndry}(r)$. This
means that the theoretical model of the integrated three-point function computed by
\refeq{iz} has to be divided by $f_{\rm bndry}(r)$. Since $\d_m(\vr)$ is Gaussian,
the only contribution to the three-point function is from the nonlinear bias term,
and so the estimated integrated three-point function is exactly given by
$\frac{\iz_{b_2}(r)}{f_{\rm bndry}(r)}$.

\refFig{mock_halo} shows the normalized integrated three-point functions of the
mock halo density field with $b_1=3$ and $b_2=1$ from 300 Gaussian realizations.
The measurements are in excellent agreement with $\frac{\iz_{b_2}(r)}{f_{\rm bndry}(r)}$.
This test gives us the confidence that our estimator is unbiased.

%%%%%%%%%%%%%%%%%%%%%%%%%%%%%%%%%%%%%%%%%%%%%%%%%%%%%%%%%%%%%%%%%%%%%%%%%%%%

\section{Effects of effective $F_2$ and $G_2$ kernels and non-local tidal bias}
\label{app:test}
In this appendix, we show how the inferred value of $b_2$ changes when extending
our baseline model for the bispectrum based on SPT at the tree level with local
bias to the model used in the analysis of ref.~\cite{gilmarin/etal:2014b}.

Their model replaces $F_2$ and $G_2$ in \refeq{treeredshift} with
``effective'' kernels, $F_2^{\rm eff}$ \cite{gilmarin/etal:2011}
and $G_2^{\rm eff}$ \cite{gilmarin/etal:2014a}, which are calibrated
to match the nonlinear matter bispectrum in of N-body simulations.
Their model also adds a non-local galaxy bias caused by tidal fields
\cite{mcdonald/roy:2009,baldauf/etal:2012,sheth/chan/scoccimarro:2012}
to $Z_2$, i.e., $Z_2\to Z_2+\frac12b_{s^2}\left[(\hat{k}_1\cdot\hat{k}_2)^2-\frac13\right]$,
where $b_{s^2}=-(4/7)(b_1-1)$. We use this model to compute the integrated
three-point function, and find $b_2$ of the mocks in real and redshift
space by performing a joint fit with the two-point function as described
in \refsec{mock_r} and \ref{sec:mock_z}.

\begin{table}
\centering
\begin{tabular}{ | c | c c | }
\hline
r-space & $b_1$ & $b_2$ \\
\hline
baseline & $1.971\pm0.076$ & $0.58\pm0.31$ \\
eff kernel & $1.973\pm0.076$ &  $0.62\pm0.31$ \\
tidal bias & $1.971\pm0.076$ &  $0.64\pm0.31$ \\
both & $1.973\pm0.076$ &  $0.68\pm0.31$ \\
\hline
\end{tabular}
~~~~~
\begin{tabular}{ | c | c c | }
\hline
z-space & $b_1$ & $b_2$ \\
\hline
baseline & $1.931\pm0.077$ & $0.54\pm0.35$ \\
eff kernel & $1.933\pm0.077$ &  $0.65\pm0.35$ \\
tidal bias & $1.932\pm0.077$ &  $0.60\pm0.35$ \\
both & $1.933\pm0.077$ &  $0.71\pm0.35$ \\
\hline
\end{tabular}
\caption{Best-fitting values of $b_1$ and $b_2$ and their uncertainties for mock catalogs,
 obtained using different models of the bispectrum in real space (left)
 and redshift space (right).}
\label{tab:test_models}
\end{table}

\refTab{test_models} summarizes the results. The ``baseline model'' refers
to the model based on SPT and local bias. The ``eff kernel'' refers to the
model with $F^{\rm eff}_2$, $G^{\rm eff}_2$, and local bias. The ``tidal
bias'' refers to the model with $F_2$, $G_2$, local bias, and tidal bias.
Finally, ``both'' refers to the model with $F^{\rm eff}_2$, $G^{\rm eff}_2$,
local bias, and tidal bias.

Both the effective kernels and the non-local tidal bias result in a larger
nonlinear bias, which is in better agreement with ref.~\cite{gilmarin/etal:2014b}.
The changes of the best-fitting nonlinear bias, however, are still within the
$1-\sigma$ uncertainties, and all the results are consistent with ref.~\cite{gilmarin/etal:2014b}.
We also calculate the goodness of the fit for all the models in both real and
redshift space by comparing the mean of the mocks and the best-fitting models,
as well as the $\chi^2$-distribution. We find that all models perform equally
well; thus, in this paper we shall primarily use the simplest model, i.e. the
SPT at the tree level with local bias for modeling the three-point function,
but also report the results for the extended models.

%%%%%%%%%%%%%%%%%%%%%%%%%%%%%%%%%%%%%%%%%%%%%%%%%%%%%%%%%%%%%%%%%%%%%%%%%%%%

\section{Comparison of $\iz(r)/\sigma_L^2$ of BOSS DR10 CMASS sample and PTHalos
mock catalogs in different redshift bins}
\label{app:iz_zevolve}
The BOSS DR10 CMASS sample and the mocks have different sets of random samples
with slightly different $\bar n(z)$, hence the properties of the observations
and the mocks may not agree well in all redshift bins. Moreover, as mentioned
in ref.~\cite{more/etal:2014}, the CMASS sample is flux-limited, and thus the
observed galaxies statistically have larger stellar masses at higher redshift
(see figure~1 in ref.~\cite{more/etal:2014}). This may cause redshift evolution
of the bias, and so the correlation functions. We shall measure $\iz(r)/\sigma_L^2$
as a function of redshift to test this.

The measurements in the subvolumes are mostly the same as introduced in \refsec{sub_quan},
except that we now measure $\alpha(z_j)$ as a function of redshift bin $z_j$,
and the average is done in the individual redshift bin. Namely,
\be
 \alpha(z_j)=\frac{\sum_{i\in z_j}w_{g,i}}{\sum_{i\in z_j}w_{r,i}}
 =\frac{w_{r,z_j}}{w_{g,z_j}}\,, ~~~~~
 \bar{g}(z_j)=\frac{1}{w_{r,z_j}}\sum_{i\in z_j}g_iw_{r,i} ~. 
\ee
This assures that $\bd(z_j)=0$ for all redshift bins.

\begin{figure}[t]
\centering
\includegraphics[width=0.325\textwidth]{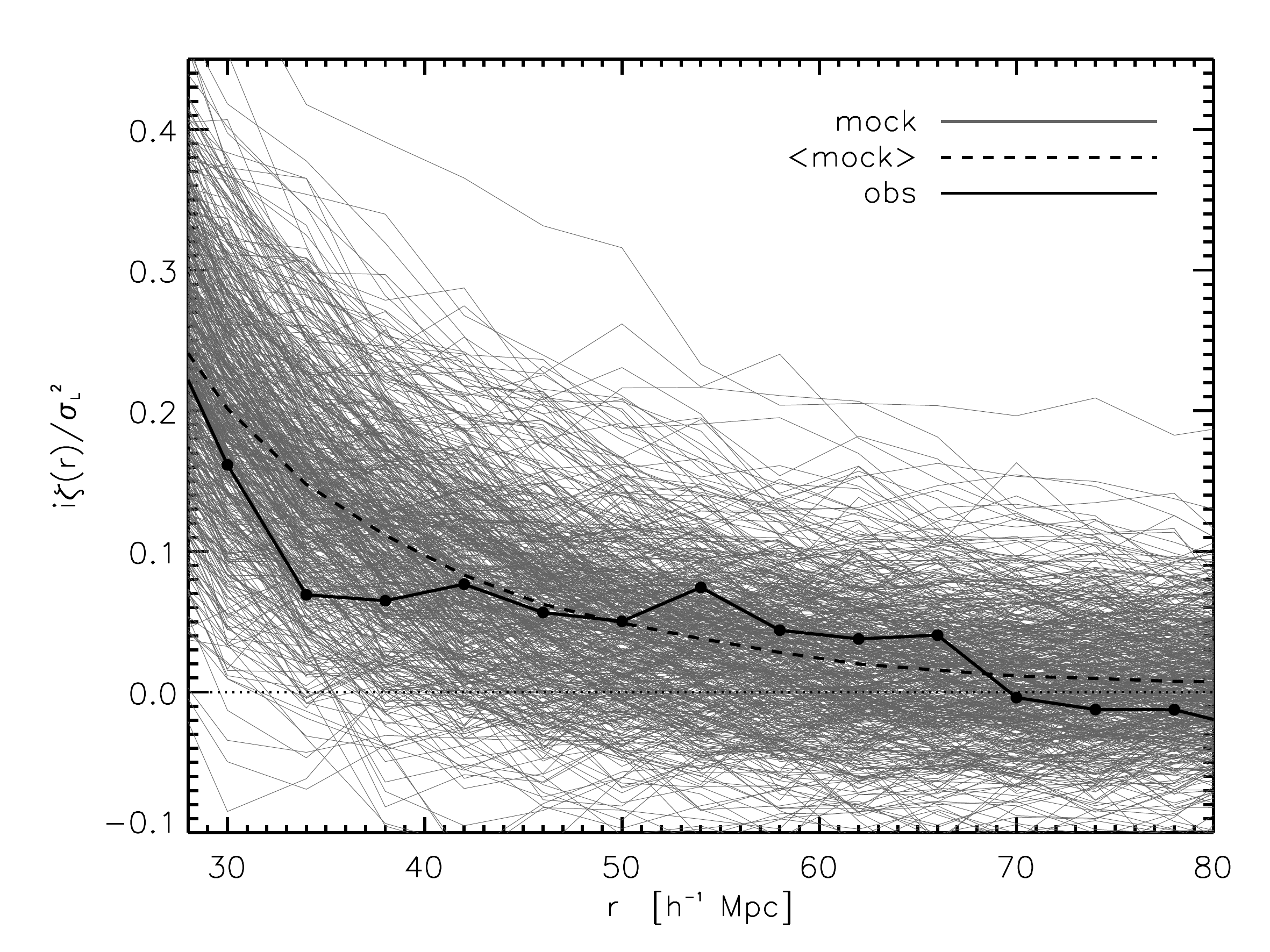}
\includegraphics[width=0.325\textwidth]{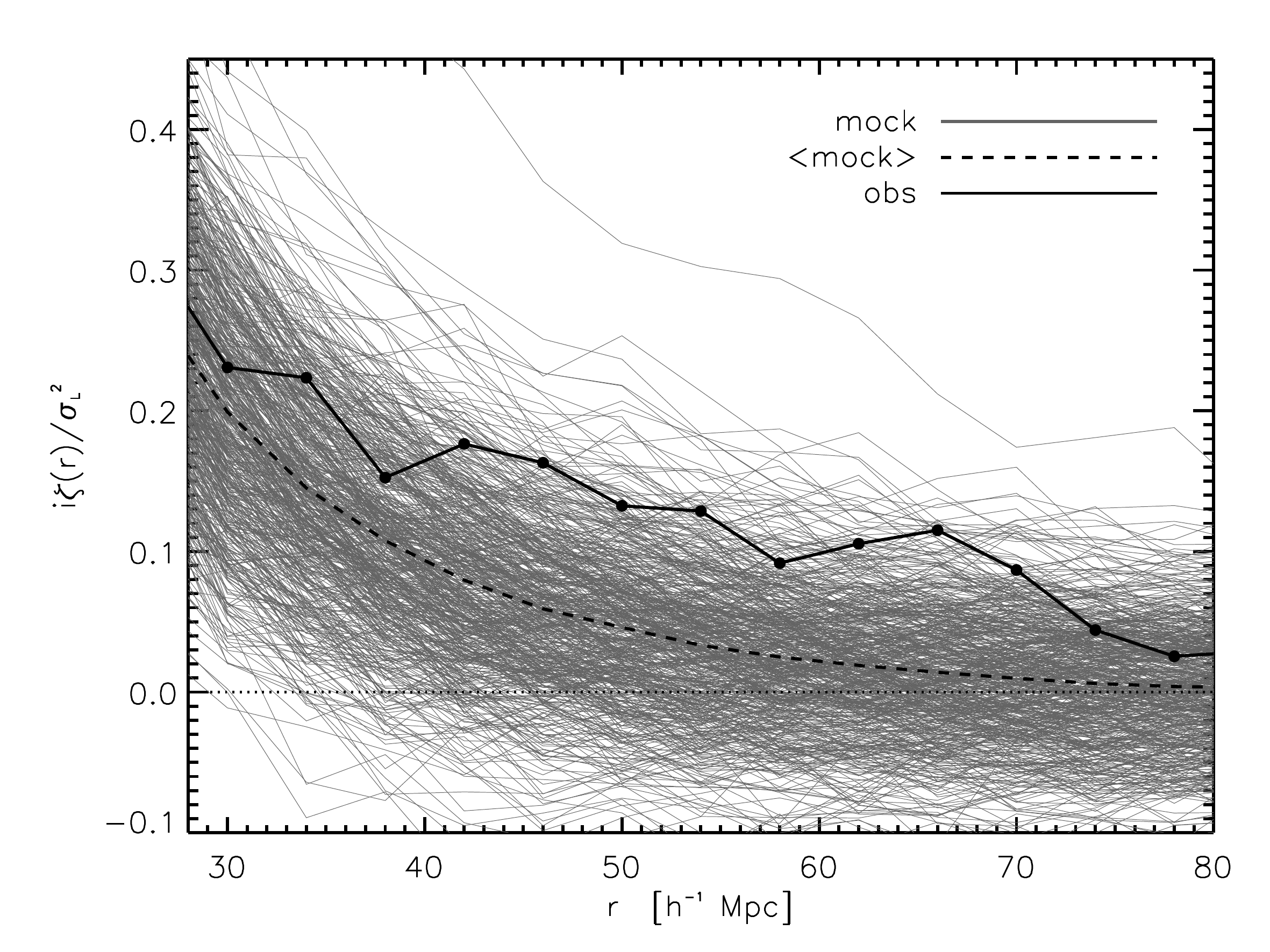}
\includegraphics[width=0.325\textwidth]{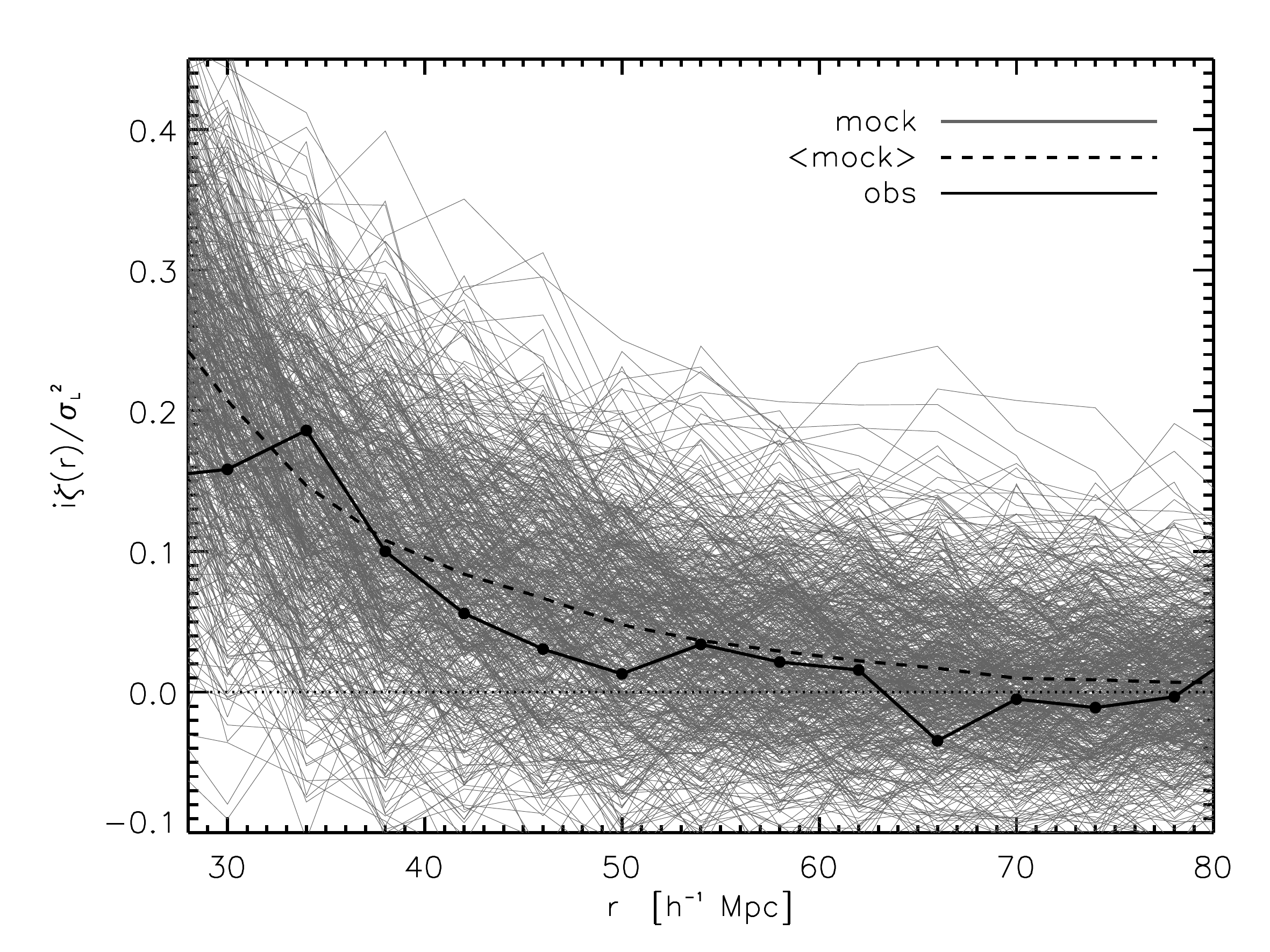}
\caption{$\iz(r)/\sigma_L^2$ of $220\hMpc$ subvolumes in different redshift bins.
The redshift bins increase from left to right, with the redshift cuts quoted in
the beginning of \refsec{mock} and \ref{sec:data}.}
\label{fig:iz_norm_zi_1}
\end{figure}

\begin{figure}[t]
\centering
\includegraphics[width=0.325\textwidth]{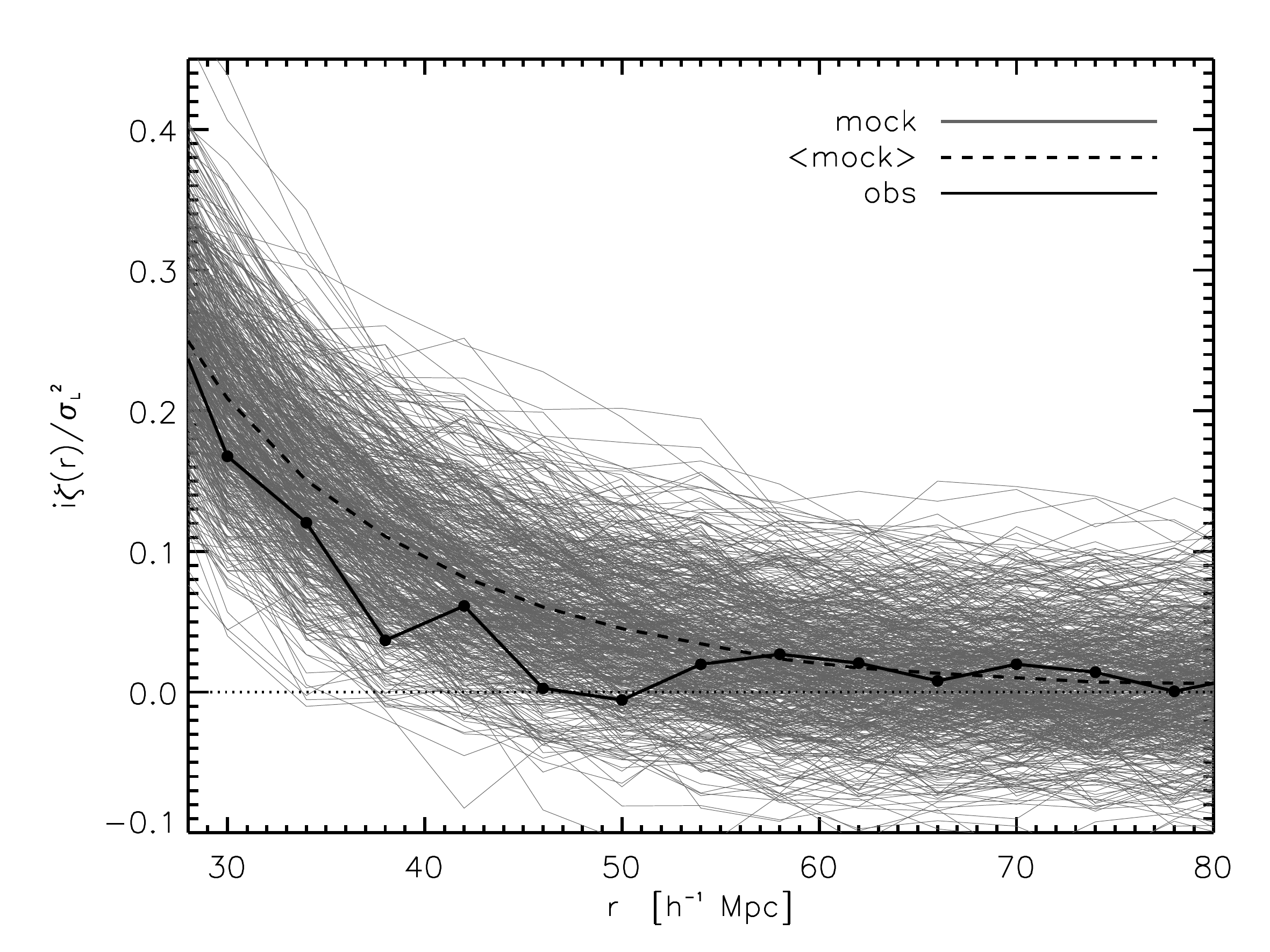}
\includegraphics[width=0.325\textwidth]{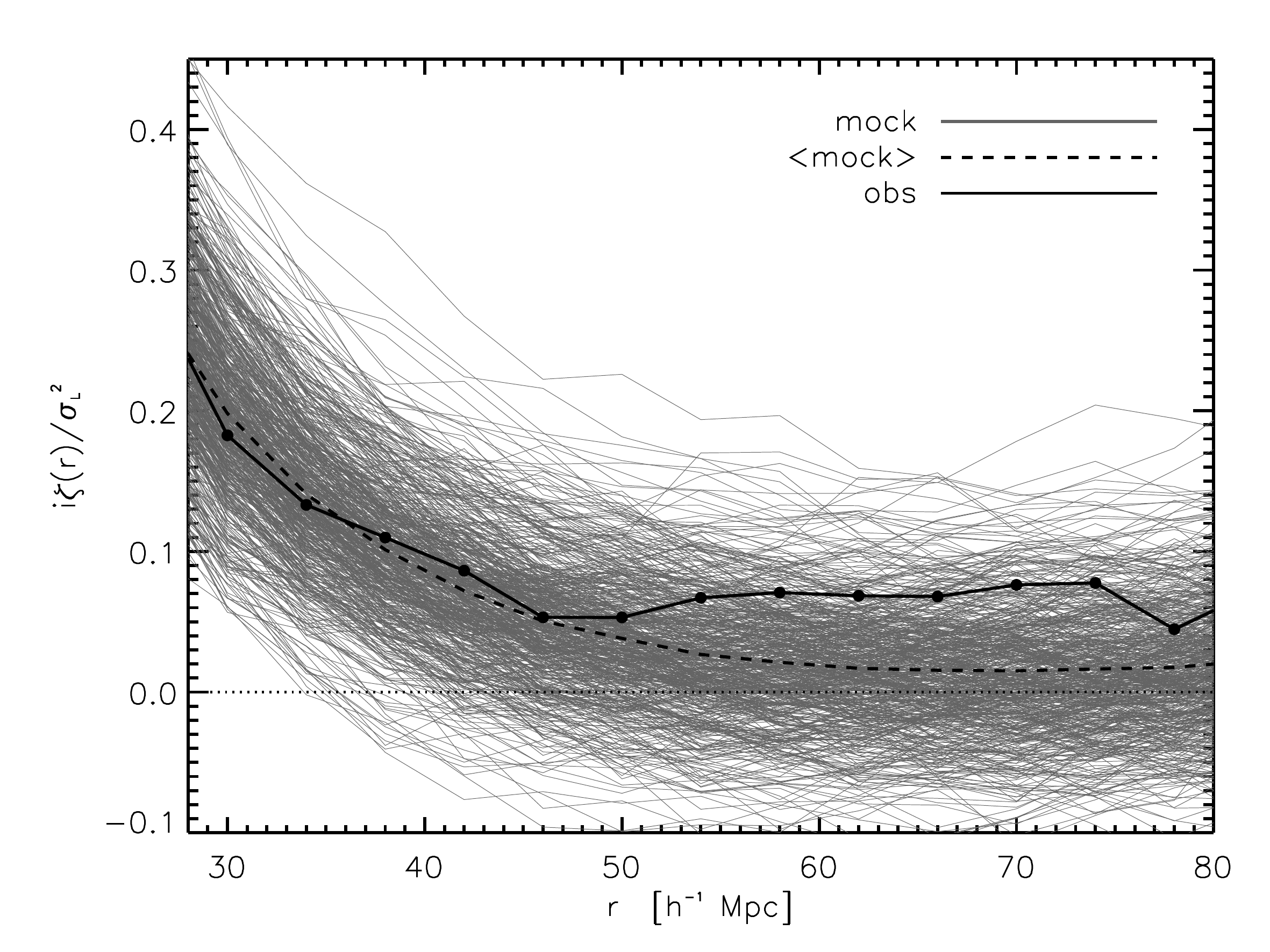}
\includegraphics[width=0.325\textwidth]{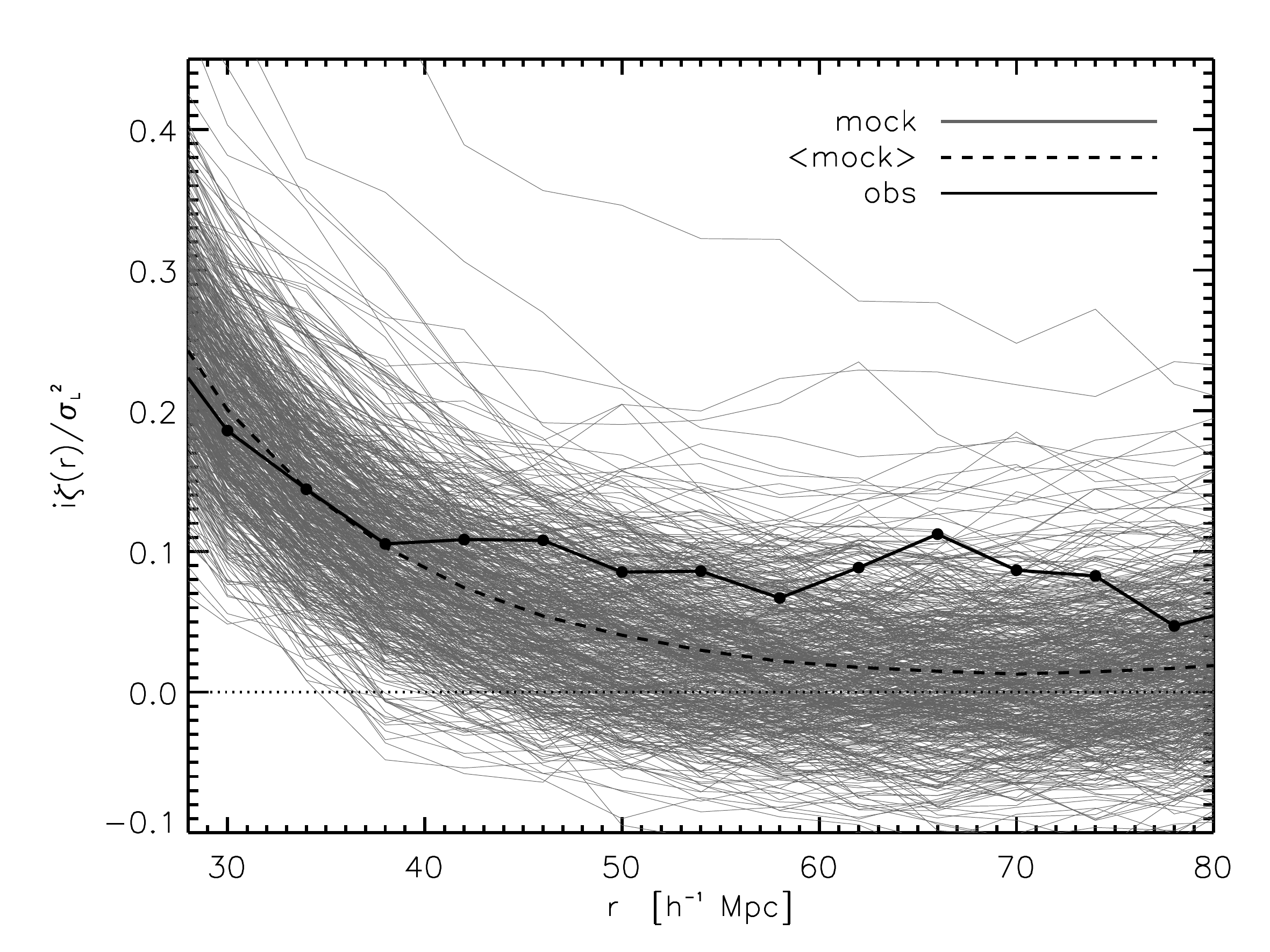}
\includegraphics[width=0.325\textwidth]{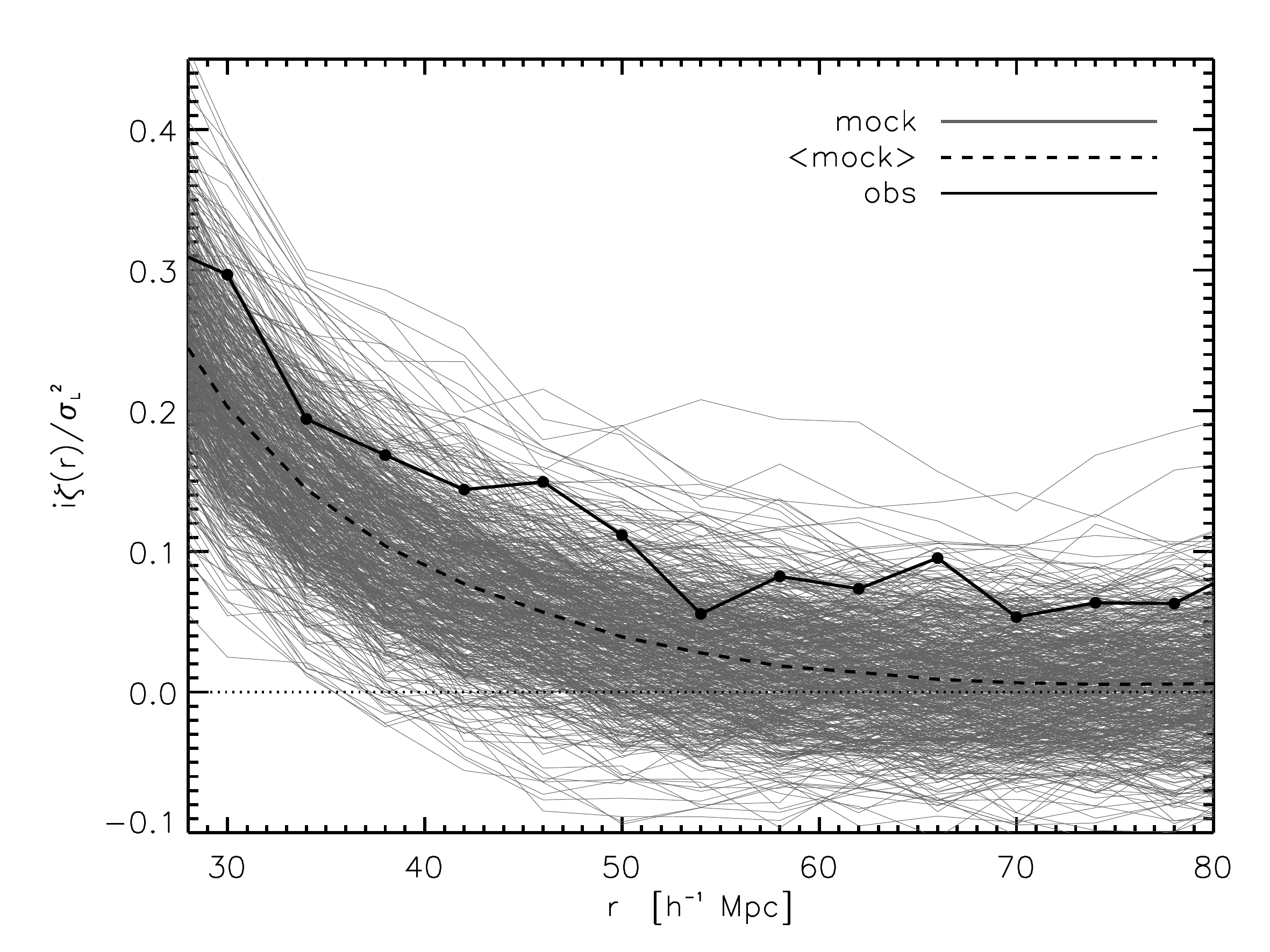}
\includegraphics[width=0.325\textwidth]{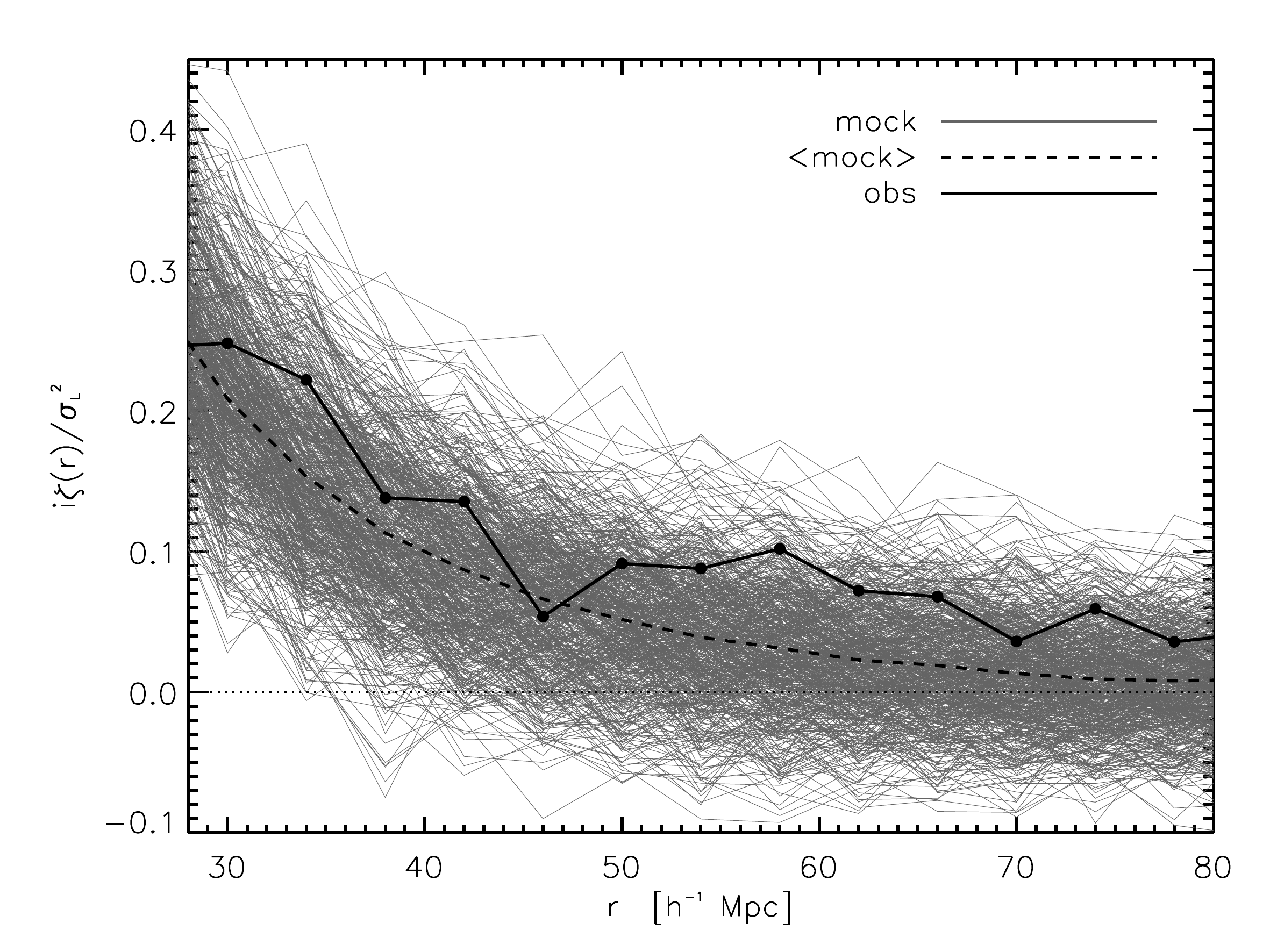}
\caption{Same as \reffig{iz_norm_zi_1}, but for $120\hMpc$ subvolumes. The redshift
bins increase from top left to bottom right.}
\label{fig:iz_norm_zi_2}
\end{figure}

\refFig{iz_norm_zi_1} and \reffig{iz_norm_zi_2} show $\iz(r)/\sigma_L^2$ at
different redshift bins for 220 and $120\hMpc$ subvolumes, respectively.
We find no clear sign that $\iz(r)/\sigma_L^2$ of the observations has
different redshift evolution relative to the mocks. Thus, it is justified
to study $\iz(r)/\sigma_L^2$ using one effective redshift for the BOSS
DR10 CMASS sample. With the upcoming DR12 sample with a larger volume,
the redshift evolution of $\iz(r)/\sigma_L^2$ can be better studied.

%%%%%%%%%%%%%%%%%%%%%%%%%%%%%%%%%%%%%%%%%%%%%%%%%%%%%%%%%%%%%%%%%%%%%%%%%%%%

\section{Fisher matrix of the position-dependent power spectrum}
\label{app:fisher}
In this appendix we demonstrate that the position-dependent correlation
function technique captures the information of the squeezed-limit bispectrum.
To this end we utilize the local type of primordial non-Gaussianity, which
peaks in the squeezed limit, and show that the full bispectrum and the
position-dependent correlation function would yield comparable constraints on $f_{\rm NL}$.
As shown in \refsec{iz_to_ib}, the integrated three-point function is the
Fourier transform of the integrated bispectrum. Here we shall compute the Fisher
matrix of the integrated bispectrum instead of the integrated three-point function.
This simplifies the computation as in this case we may assume the covariance matrix to be diagonal (in the Gaussian limit).

As the main goal of this appendix is to show that the information of the
squeezed-limit bispectrum is captured by the position-dependent correlation
function technique, we shall use the simplest model for the galaxy bispectrum
in the presence of local-type primordial non-Gaussianity and redshift-space
distortion. Thus, the galaxy bispectrum is given by
\be
 B_g(\vk_1,\vk_2,vk_3)=K_b\left[b_1^3B_{\rm SPT}(\vk_1,\vk_2,\vk_3)
 +b_1^2b_2B_{b_2}(\vk_1,\vk_2,\vk_3)+b_1^3f_{\rm NL}B_{f_{\rm NL}}(\vk_1,\vk_2,\vk_3)\right] ~,
\ee
where $K_b=1+\frac23\beta+\frac19\beta^2$ is the Kaiser factor of the bispectrum
and $\beta=f/b_1$ \cite{kaiser:1987}, and $f_{\rm NL}$ parametrizes the amplitude
of the primordial non-Gaussianity. The exact forms of the bispectra are
\ba
 B_{\rm SPT}(\vk_1,\vk_2,\vk_3)\:&=2F_2(\vk_1,\vk_2)P_l(k_1,a)P_l(k_2,a)+2~{\rm cyclic} \vs
 B_{b_2}(\vk_1,\vk_2,\vk_3)\:&=P_l(k_1,a)P_2(k_2,a)+2~{\rm cyclic} \vs
 B_{f_{\rm NL}}(\vk_1,\vk_2,\vk_3)\:&=2M(k_1,a)M(k_2,a)M(k_3,a)
 \left[P_{\Phi}(k_1)P_{\Phi}(k_2)+2~{\rm cyclic}\right] ~,
\label{eq:ib_fnl}
\ea
where $a$ is the scale factor, $M(k,a)=\frac23\frac{D(a)}{H_0^2\Omega_m}k^2T(k)$
and $T(k)$ is the transfer function, and $P_{\Phi}(k)$ is the power spectrum of
the scalar potential. $B_{f_{\rm NL}}$ is produced by the local-type non-Gaussianity
in the primordial scalar potential, $\Phi(r)=\phi(r)+f_{\rm NL}\left[\phi^2(r)-\langle\phi^2(r)\rangle\right]$
where $\phi(r)$ follows the Gaussian statistics \cite{komatsu/spergel:2001}. Note
that we ignore the effect of the scale-dependent bias due to the local-type primordial
non-Gaussianity \cite{dalal/etal:2007,matarrese/verde:2008,slosar/etal:2008}, and the
more complete model can be found in \cite{baldauf/seljak/senatore:2010,tasinato/etal:2013}.

The Fisher matrix of the reduced bispectrum $Q(k_1,k_2,k_3)=B(k_1,k_2,k_3)/[P(k_1)P(k_2)+2~{\rm cyclic}]$
is given by \cite{sefusatti/komatsu:2007}
\be
 F_{Q,\alpha\beta}=\sum_{k_1,k_2,k_3\le k_{\rm max}}\frac{\partial Q(k_1,k_2,k_3)}{\partial p_{\alpha}}
 \frac{\partial Q(k_1,k_2,k_3)}{\partial p_{\beta}}\frac1{\Delta Q^2(k_1,k_2,k_3)} ~,
\label{eq:fisher_q}
\ee
where $(k_1,k_2,k_3)$ have to form a triangle, $p_{\alpha}\in[b_1,b_2,f_{\rm NL}]$
are the parameters, and $\Delta Q^2(k_1,k_2,k_3)$ is the variance of the reduced
bispectrum. Similarly, the Fisher matrix of the normalized integrated bispectrum
$ib(k)=iB(k)/[P(k)\sigma_L^2]$ is given by
\be
 F_{ib,\alpha\beta}=\sum_L\sum_{k\le k_{\rm max}}\frac{\partial ib(k)}{\partial p_{\alpha}}
 \frac{\partial ib(k)}{\partial p_{\beta}}\frac1{\Delta ib^2(k)} ~,
\label{eq:fisher_ib}
\ee
where $L$ is the subvolume size and $\Delta ib^2(k)$ is the variance of the normalized
integrated bispectrum. Note that we assume the Gaussian limit in \refeqs{fisher_q}{fisher_ib},
hence only the diagonal elements of $Q$ and $ib$ exist. Namely, for $Q$, it requires
$k_i=k_i'$ for $\langle Q(k_1,k_2,k_3)Q(k_1',k_2',k_3')\rangle$; similarly, for $ib$,
it requires $L=L'$ and $k=k'$. This approximation holds well at high redshift, but
would break down at low redshift due to the nonlinear evolution.

To compute the variances, we assume that the dominant component is from the bispectrum
in the numerator (instead of the normalization in the denominator). We thus have
\ba
  \Delta Q^2(k_1,k_2,k_3)\approx\:&\frac{\pi s_{123}}{k_1k_2k_3}
 \frac{[P_z(k_1)+P_{\rm shot}][P_z(k_2)+P_{\rm shot}][P_z(k_3)+P_{\rm shot}]}
 {[P_z(k_1)P_z(k_2)+{\rm 2~cyclic}]^2} ~, \vs
 \Delta ib^2(k)\approx\:&\frac{V_L}{V_rN_{kL}}
 \frac{[\sigma_{L,z}^2+P_{\rm shot}/V_L][P_z(k)+P_{\rm shot}]^2}{\sigma_{L,z}^4P_z^2(k)} ~,
\ea
where $s_{123}=6$, 2, 1 for equilateral, isosceles, and general triangles, respectively,
$P_z(k)=K_pb_1^2P_l(k)$ is the redshift-space power spectrum, $P_{\rm shot}$ is the shot
noise of the power spectrum, $N_{kL}$ is the number of independent Fourier modes in the
subvolume, and $\sigma_{L,z}^2=K_pb_1^2\sigma_{L,l}^2$.

\begin{figure}[t]
\centering
\includegraphics[width=0.6\textwidth]{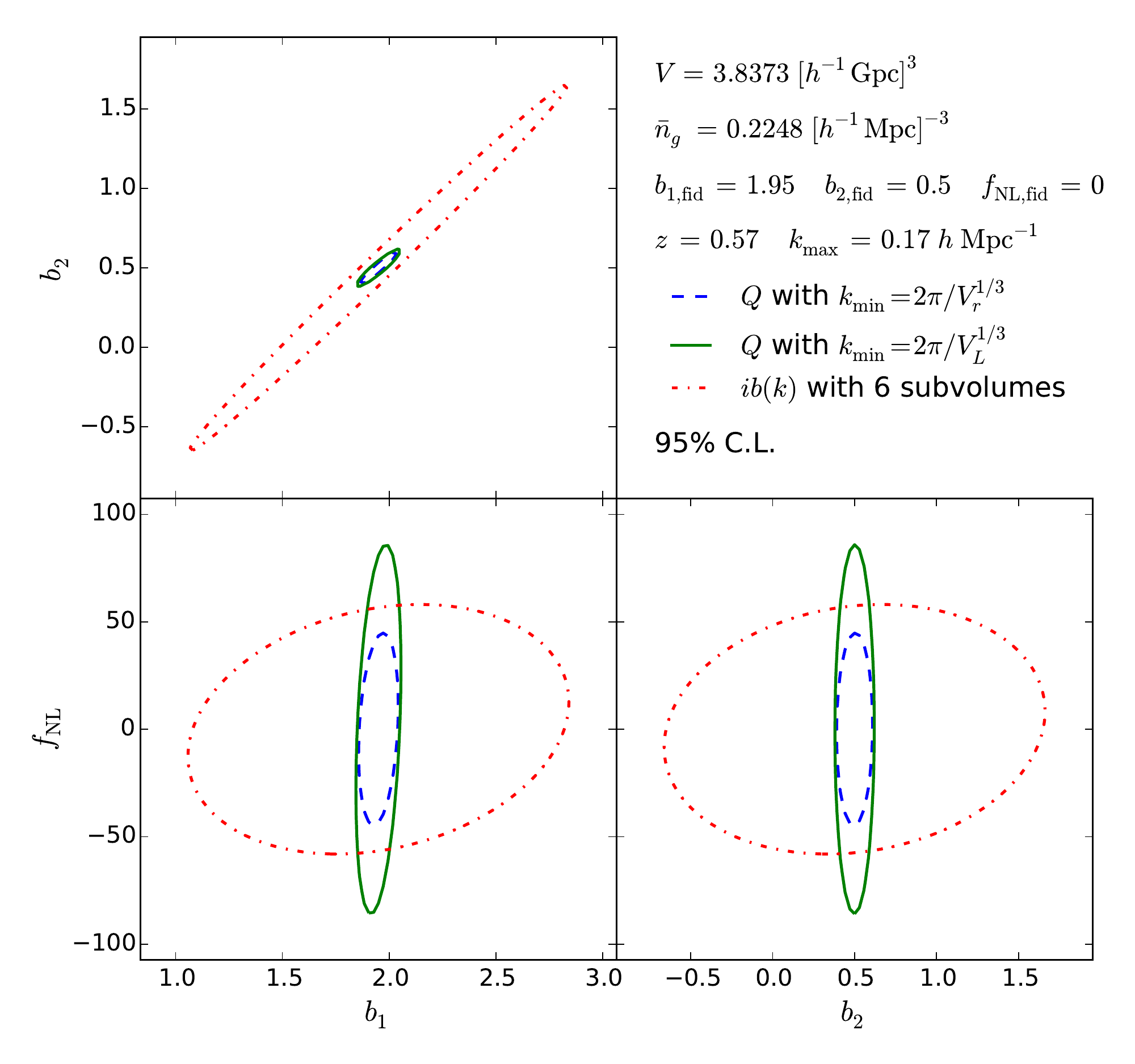}
\caption{Two-dimensional joint 95\% C.L. constraints on galaxy biases and primordial
non-Gaussianity for BOSS full survey. The survey parameters are in the top right panel.
The top-left, bottom-left, and bottom-right panels show the joint constraints on
$(b_1,b_2)$, $(b_1,f_{\rm NL})$, and $(b_2,f_{\rm NL})$ marginalized over $f_{\rm NL}$,
$b_2$, and $b_1$, respectively. The blue dashed, green solid, and red dot-dashed line
are for reduced bispectrum with $k_{\rm min}=2\pi/V_r^{1/3}$, reduced bispectrum with
$k_{\rm min}=2\pi/V_L^{1/3}$ where $V_L$ is the largest subvolume, and normalized
integrated bispectrum for six sizes of subvolumes (100 to $600\hMpc$ with an increment
of $100\hMpc$), respectively.}
\label{fig:fisher}
\end{figure}

\refFig{fisher} shows the two-dimensional joint 95\% C.L. constraints on galaxy biases
and primordial non-Gaussianity for the full BOSS survey. One finds that as long as
$k_{\rm min}$ is set to be the fundamental frequency of the largest subvolume, the
integrated bispectrum technique gives the similar constraint on $f_{\rm NL}$ compared
to the full bispectrum analysis.\footnote{Note that the numerical results are sensitive
to the choices of $k_{\rm min}$ and $k_{\rm max}$ because we count the Fourier modes
in this range. For different lines, though $k_{\rm max}$ is set to be the same, in
practice we stop counting Fourier modes if $k>k_{\rm max}$. Therefore, they have different
``true'' $k_{\rm max}$, and the contour area would be affected. This explain why the green
solid line has slightly larger area than that of the red dot-dashed line in the bottom panels.}
This means that the integrated bispectrum is sensitive to the bispectrum in the
squeezed-limit configurations. On the other hand, the top-left panel shows the strong
degeneracy between $b_1$ and $b_2$ for the integrated bispectrum. As discussed in
\refsec{int_xi}, in the squeezed limit $B_{\rm SPT}$ and $B_{b_2}$ has the same scale-dependency
if the power spectrum is pure power-law. Therefore, if only the squeezed-limit bispectrum
is used, such as the integrated bispectrum technique, then it is challenging to break
the degeneracy.

We also find that while both techniques give similar constraints on $f_{\rm NL}$, the
number of counted Fourier modes differ dramatically. Specifically, the full bispectrum
using $k_{\rm min}=2\pi/V_r^{1/3}$ and $2\pi/V_L^{1/3}$ counts 7113 and 6730 triangles,
respectively, while the integrated bispectrum counts only 54 Fourier modes. This means
that many of the triangles contain information other than the squeezed configurations\footnote{This
explains why the full bispectrum is more powerful for breaking the degeneracy between
$b_1$ and $b_2$.} and thus the integrated bispectrum can be regarded as an efficient
approach to extract the information of the squeezed-limit bispectrum, such as the
local-type primordial non-Gaussianity.

%%%%%%%%%%%%%%%%%%%%%%%%%%%%%%%%%%%%%%%%%%%%%%%%%%%%%%%%%%%%%%%%%%%%%%%%%%%%

\bibliography{references}
\end{document}